\documentclass[
superscriptaddress,
preprint,
prd,tightenlines,showpacs,nofootinbib,
eqsecnum,
amsfonts,amsmath,amssymb]{revtex4-1}

\usepackage{bm}
\usepackage{graphicx} 
\usepackage{hyperref}
\usepackage{mathrsfs}
\usepackage{multirow}

\newcommand{\nn}{\nonumber}

\newcommand{\ov}[1]{\overline{#1}}
\newcommand{\ovbf}[1]{\overline{\mathbf{#1}}}
\newcommand{\ovbm}[1]{\overline{\bm{#1}}}
\newcommand{\ud}{\mathrm{d}}
\newcommand{\uD}{\mathrm{D}}

\newcommand{\calO}{\mathcal{O}}

\newcommand{\ph}[1]{\phantom{#1}}

\usepackage{ulem}
\usepackage[usenames]{color}

\allowdisplaybreaks

\begin{document}

\title{Next-to-next-to-leading order spin-orbit effects in the \\near-zone
  metric and precession equations of compact binaries}

\author{Alejandro \textsc{Boh\'{e}}}\email{alejandro.bohe@uib.es}
\affiliation{Departament de F\'isica, Universitat de les Illes Balears, Crta.
  Valldemossa km 7.5, E-07122 Palma, Spain}

\author{Sylvain \textsc{Marsat}}\email{marsat@iap.fr}
\affiliation{$\mathcal{G}\mathbb{R}\varepsilon{\mathbb{C}}\mathcal{O}$
  Institut d'Astrophysique de Paris --- UMR 7095 du CNRS, \ Universit\'e
  Pierre \& Marie Curie, 98\textsuperscript{bis} boulevard Arago, F-75014 Paris,
  France}

\author{Guillaume \textsc{Faye}}\email{faye@iap.fr}
\affiliation{$\mathcal{G}\mathbb{R}\varepsilon{\mathbb{C}}\mathcal{O}$
  Institut d'Astrophysique de Paris --- UMR 7095 du CNRS, \ Universit\'e
  Pierre \& Marie Curie, 98\textsuperscript{bis} boulevard Arago, F-75014 Paris,
  France}

\author{Luc \textsc{Blanchet}}\email{blanchet@iap.fr}
\affiliation{$\mathcal{G}\mathbb{R}\varepsilon{\mathbb{C}}\mathcal{O}$
  Institut d'Astrophysique de Paris --- UMR 7095 du CNRS, \ Universit\'e
  Pierre \& Marie Curie, 98\textsuperscript{bis} boulevard Arago, F-75014 Paris,
  France}

\date{\today}

\begin{abstract}
  We extend our previous work devoted to the computation of the
  next-to-next-to-leading order spin-orbit correction (corresponding
  to 3.5PN order) in the equations of motion of spinning compact
  binaries, by: (i) Deriving the corresponding spin-orbit terms in the
  evolution equations for the spins, the conserved integrals of the
  motion and the metric regularized at the location of the particles
  (obtaining also the metric all-over the near zone but with some
    lower precision); (ii) Performing the orbital reduction of the
  precession equations, near-zone metric and conserved integrals to
  the center-of-mass frame and then further assuming quasi-circular
  orbits (neglecting gravitational radiation reaction). The results
  are systematically expressed in terms of the spin variables with
  conserved Euclidean norm instead of the original antisymmetric spin
  tensors of the pole-dipole formalism. This work paves the way to the
  future computation of the next-to-next-to-leading order spin-orbit
  terms in the gravitational-wave phasing of spinning compact
  binaries.
\end{abstract}


\maketitle


\section{Introduction}
\label{Introduction}

Coalescing binaries of compact objects --- neutron stars and/or black
holes --- are one of the most promising sources for the detection of
gravitational waves (GW) with the advanced versions of the
ground-based interferometers LIGO \cite{Abbott2009,LIGOwebsite} and
VIRGO \cite{Accadia2012a,VIRGOwebsite}, with GEO-HF \cite{Grote2008}
which will start taking data in the next years, and with the cryogenic
detector KAGRA joining the network in a relatively near future
\cite{Kuroda2010}. Further ahead, LISA-type space-based
interferometers \cite{Amaro-Seoane2012,eLISAwebsite}, which will
significantly increase the accessible region of parameter space, will
be able to detect supermassive black-hole binaries with a very high
signal-to-noise ratio. Successfully extracting the possibly very weak signal
from the noise or estimating the parameters of the source with good
accuracy and precision can be achieved by using matched filtering
techniques, provided that the waveform is priorly correctly modeled
(see \textit{e.g.} Refs.~\cite{Cutler1993, Cutler1994}). The
post-Newtonian (PN) approximation scheme enables the computation of
precise/accurate templates for the inspiralling phase of the
coalescence of compact binaries \cite{Blanchet2006a}.

For non-spinning objects, both the dynamics of the system and the
waveform phase have been derived up to the 3.5PN order
\cite{Blanchet2004a,Arun2004,Blanchet2002}.\footnote{As usual the
  $n$PN order includes corrections up to the relative order
  $1/c^{2n}$ in a power expansion in the inverse of the speed of
  light.} In the last years, an important effort (motivated by
astrophysical observations
\cite{Abramowicz2001,Strohmayer2001b,McClintock2006,Gou2011,Nowak2012})
has been undertaken to extend these results to the spinning case, up
to the same accuracy. In the present work, we investigate the
higher-order binary dynamics focusing on the \textit{spin-orbit}
effects, which are linear in the spin parameter and numerically the
most important to be taken into account at leading order. Here by
spin, we mean the intrinsic (classical) angular momentum $S$ of the
individual compact body, rescaled in our convention by a factor $c$ as
$S\equiv c S_{\text{true}}=G m_{\text{body}}^2 \chi$, where $\chi$ is
the dimensionless spin parameter equal to $1$ for maximally spinning
objects, so that $S$ appears to be formally of Newtonian order and all
powers of $1/c$ are kept explicitly. Adopting this power counting, the
leading order spin-orbit and spin-spin contributions to the dynamics
appear at 1.5PN and 2PN respectively
\cite{Barker1975b,BarkerOConnell2,KWW93,Kidder1995,Goldberger2006,Porto2006},
while the next-to-leading corrections are of 2.5PN
\cite{Owen1998,Tagoshi2001,Faye2006,Damour2008a,Levi2010a,Porto2010a}
and 3PN order \cite{Hergt2010,Steinhoff2008,Hartung2011a,Porto2008a,Porto2008b}
respectively. Note that the spin-spin couplings between different
spins (1 and 2) are actually known at 4PN order
\cite{Levi2012a,Hartung2011c}.

In a recent work \cite{Marsat2012},\footnote{Hereafter this work will
  be referred to as Paper~I.} we computed the next-to-next-to leading
order spin-orbit correction at order 3.5PN in the equations of
motion. We used a direct post-Newtonian iteration of the Einstein
field equations in harmonic coordinates and proved the equivalence of
our result with the one obtained in Ref.~\cite{Hartung2011} using a
Hamiltonian approach in ADM coordinates. We have further computed the
associated conserved energy, verified the manifest Lorentz invariance
of the equations of motion, and found agreement in the test mass limit
with the motion of a spinless test particle around a Kerr black hole,
as well as that of a spinning test particle around a Schwarzschild
black hole. Our calculation was based on the description of the two
compact objects as spinning point particles within the framework of
the pole-dipole effective model. We refer to Section II of Paper~I and
references therein for a detailed presentation of the formalism on
which the present work also relies. Our purpose here, building on
Paper~I, is threefold:
\begin{enumerate}
\item Deriving the next-to-next-to-leading order spin-orbit
  corrections to the evolution equations for the spins, the conserved
  integrals of motion and the metric regularized at the location of
  the particles. We also obtain an expression for the spin-orbit part
  of the metric valid all over the near-zone up to the highest
  available order. Knowing explicitly the spin contributions in the
  near-zone metric is important for applications such as the numerical
  study of the accretion disk or jet dynamics around spinning back
  hole binaries \cite{GNYC12}, and the on-going comparison
  between PN predictions and the numerical calculation of the self
  force acting on a particle orbiting a spinning black hole
  \cite{BBL12,FLS12}.
\item Expressing the results in terms of spin variables with conserved
  Euclidean norm instead of the original antisymmetric tensors of the
  pole-dipole formalism as used in Paper~I. This choice simplifies the
  dynamics since the spin evolution equations take the form of
  ordinary precession equations, and is motivated by our final goal,
  namely to compute the next-to-next-to-leading spin-orbit effects in
  the GW orbital phase. The evolution of the GW phase will be indeed
  obtained via an energy balance argument requiring specifically the
  use of the conserved-norm spins.
\item Performing the reduction of our results to the center-of-mass
  frame and to the case of quasi-circular orbits. This is of
  particular interest for data analysis purposes since, generally,
  orbits are circularized by emission of gravitational
  radiation by the time the binary enters the observational
  band. Notably, we obtain the 3.5PN expression including spin-orbit
  terms of the binary's energy as a function of the orbital frequency,
  which is one of the main ingredients entering the construction of
  the various approximants used for the evolution of the GW phase.
\end{enumerate}

The paper is organized as follows. In Section \ref{sec:spinvar}, we
present the construction of the conserved-norm spin variable that we
use throughout the paper as well as the associated precession equation
including the next-to-next-to leading spin-orbit effects, written in a
general frame. The other general-frame expressions being too long to
be shown here, we directly move on to the presentation of our results
in the center-of-mass frame in Section \ref{sec:CM}. The reduction to
quasi-circular orbits is performed in Section \ref{sec:CO}, and the
Section \ref{sec:conclusion} contains our conclusions. In Appendix
\ref{app:CompADM} we compute the correspondence between our spin
variables and those of Ref.~\cite{Hartung2011}, whereas Appendix
\ref{app:Explicitspinrelation} gives the explicit relation between our
spin variable and the antisymmetric spin tensor of Paper~I. Appendix
\ref{appendix:CMreduction} gives the general frame expression of the
center-of-mass position at next-to-next-to leading spin-orbit
order. We relagate in Appendix \ref{app:angmom} the lengthy
  expression of the total angular momentum. In Appendix \ref{3PNgauge}
  we show that spin-orbit terms at 3PN order are pure gauge.

Most of our computations were achieved by means of the package xAct,
which handles symbolic tensor calculus within the scientific software
Mathematica{\footnotesize \textregistered} \cite{xtensor}.

\section{Precession equations of compact binaries}
\label{sec:spinvar}

\subsection{Definition of the constant magnitude spin}

While in Paper~I we found most convenient to work with the space
components of the spin tensor, in the present article we shall
introduce new spin variables, denoted $S_{a}$ (with $a=1,2,3$), which
are designed to have a conserved Euclidean norm. Using conserved-norm
spin vector variables is indeed the most natural choice when
  considering the dynamics reduced to the center-of-mass frame or to
circular orbits. Their evolution equations reduce, by construction, to
ordinary precession equations, and these variables are important when
studying the gravitational waves emitted by a quasi-circular binary
because they are secularly constant \cite{Will05,Blanchet2006}.

For convenience, we change our notations for the spin variables with
respect to Paper~I. For the spin tensor variable of Paper~I we now use
the notation:
\begin{equation}
	\tilde{S}^{\mu\nu} \equiv S^{\mu\nu}_\text{MBFB} \;.
\end{equation}
Let us recall from Section~II in Paper~I that this variable satisfies
the covariant spin supplementary condition (SSC), namely
$\tilde{S}^{\mu\nu}p_{\mu}=0$ where $p_\mu$ denotes the linear
four-momentum of the particle. The covariant spin vector
$\tilde{S}_{\mu}$ associated with the spin tensor is therefore defined
by\footnote{For convenience in this Subsection we pose $c=1$.}
\begin{equation}\label{eq:DefSmu}
  \tilde{S}^{\mu\nu} \equiv -\frac{1}{\sqrt{-g}} \,
  \varepsilon^{\mu\nu\rho\sigma}\frac{p_{\rho}}{m}
  \tilde{S}_{\sigma} \;,
\end{equation}
where $\varepsilon^{\mu\nu\rho\sigma}$ denotes the Levi-Civita symbol;
we know that both $p_\mu p^\mu = - m^2$ and
$\tilde{S}_{\mu}\tilde{S}^{\mu} = s^{2}$ are conserved along the
trajectory: $m = \mathrm{const}$ and $s = \mathrm{const}$. Working at
linear order in the spins, the linear momentum agrees with the
normalized four velocity, $p_\mu = m \,u_\mu + \calO(S^{2})$, and we
simply get that:
\begin{equation}\label{parallel}
  \frac{\uD\tilde{S}_{\mu}}{\ud\tau} = \calO(S^{2}) \; ,
\end{equation}
where $\uD/\ud\tau \equiv u^{\nu}\nabla_{\nu}$. Thus the spin covector
is parallel transported at linear order in spin. We can also impose
that the spin should be purely spatial for the comoving observer:
\begin{equation}
	\tilde{S}_{\mu}u^{\mu} = \calO(S^{2})\; .
\end{equation}
From now on, we shall omit writing the $\calO(S^{2})$ remainders.

A standard, general procedure to define a constant-norm spin vector
consists in projecting $\tilde{S}_\mu$ onto some orthonormal tetrad
$e_\alpha^{\phantom{\alpha}\mu}$, \textit{i.e.} a tetrad that satisfies
$g_{\mu\nu}e_\alpha^{\phantom{\alpha}\mu}e_\beta^{\phantom{\beta}\nu}=\eta_{\alpha\beta}$,
which leads to the four scalar components
\begin{equation}\label{constantspin0}
	S_\alpha = e_\alpha^{\phantom{\alpha}\mu} \tilde{S}_{\mu}\; .
\end{equation}
If we choose for the time-like tetrad vector the four-velocity itself,
$e_0^{\phantom{0}\mu} = u^\mu$, the time component tetrad projection
$S_{0}$ vanishes because of the previous orthogonality condition
$\tilde{S_{\mu}}u^{\mu}=0$, and the spatial components $S_a$ (with
$a=1,2,3$) define a constant-norm spin vector. Indeed we have seen
that $\tilde{S}_{\mu}\tilde{S}^{\mu} = s^{2} = \mathrm{const}$ is
conserved along the trajectory. Because $\tilde{S_{\mu}}u^{\mu}=0$ we
can rewrite this as $\gamma^{\mu\nu}\tilde{S}_{\mu}\tilde{S}_{\nu} =
s^{2}$, in which we have introduced the projector
$\gamma^{\mu\nu}=g^{\mu\nu}+u^\mu u^\nu$ onto the spatial hypersurface
orthogonal to $u^\mu$. Now, from the orthonormality of the tetrad
and from our choice $e_0^{\phantom{0}\mu} = u^\mu$, we have
\begin{equation}\label{projecteur}
	\gamma^{\mu\nu}=\delta^{ab}e_a^{\phantom{a}\mu}e_b^{\phantom{\beta}\nu}\; .
\end{equation}
Therefore the conservation law
$\gamma^{\mu\nu}\tilde{S}_{\mu}\tilde{S}_{\nu} = s^{2}$ becomes
\begin{equation}\label{orthog}
	\delta^{ab}S_{a}S_{b} = s^{2}\; ,
\end{equation}
which is the relation defining a Euclidean constant-magnitude spin
variable $S_a$.

However, the choice of the spin variable $S_a$ is still somewhat
arbitrary, since a rotation of the tetrad vectors can always be
performed. Here, in order to fix it, we shall achieve a construction
equivalent to that of Ref.~\cite{Damour2008a}, which presents the
advantage of providing a unique determination of these redefined
variables in a given gauge. This results in a definition which differs
from the one adopted at the previous order in
Refs.~\cite{Faye2006,Blanchet2006}. In Appendix~\ref{app:CompADM} we
shall study the correspondence between our constant magnitude spin
variable and the ADM one, used in
Refs.~\cite{Damour2008a,Hartung2011}.

To uniquely fix the tetrad vectors we proceed in the following
way. Consider the spatial \textit{covariant} components of the
projector tensor $\gamma_{\mu\nu}$, namely $\gamma_{ij} = g_{ij} + u_i
u_j$, which as we see from Eq.~\eqref{projecteur} can also be written
in terms of space tetrad vectors as
$\gamma_{ij}=\delta^{ab}e_{ai}e_{bj}$, where the ``mixed-components''
tetrad vectors read $e_{ai}=e_a^{\phantom{a}\mu}\gamma_{i\mu}$. The
problem is thus to find a prescription for defining the $3\times
  3$ matrix $e_{ai}$ starting from the $3\times 3$ matrix
$\gamma_{ij}$. Clearly, if $e_{ai}$ is symmetric in its two indices,
namely $e_{ai}=e_{ia}$ (in which we are actually exchanging a
covariant spatial index $i$ with the Lorentz spatial index $a$), then
the matrix $e_{ai}$ can be viewed as the \textit{square root} of the
matrix $\gamma_{ij}$. Now, as pointed out in Ref.~\cite{Damour2008a},
a positive-definite symmetric matrix such as $\gamma_{ij}$ has a
\emph{unique} symmetric positive-definite square root $e_{ai}$, which
therefore satisfies $\gamma_{ij}=\delta^{ab}e_{ia}e_{bj}$. Thanks to
this
lemma\footnote{The existence of a symmetric positive-definite square
  root of a symmetric positive-definite matrix comes from the fact
  that the latter can be diagonalized and has only positive
  eigenvalues. The unicity is easily proven after remarking that the
  original matrix and its square root commute and therefore can be
  \textit{simultaneously} diagonalized.}  we have a mean to
define the tetrad in a ``canonical'' way, by adopting for $e_{ai}$ the
unique symmetric positive-definite square root of $\gamma_{ij}$. We
find
\begin{equation}\label{tetrad}
	e_a^{\phantom{a}\mu} = \left(\gamma^{\mu i}-\gamma^{\mu 0}v^i\right)e_{ai}  \; ,
\end{equation}
where we have used $e_{a0}=-v^ie_{ai}$ with $v^i=u^i/u^0$ denoting the
ordinary coordinate velocity. This, together with
$e_0^{\phantom{0}\mu}=u^\mu$, defines completely and uniquely the
tetrad and therefore the constant-magnitude spin variable
\begin{equation}\label{constantspin}
  S_a = e_a^{\phantom{a}\mu} \tilde{S}_{\mu} = \left(e_a^{\phantom{a}i} 
    - v^i e_a^{\phantom{a}0}\right) \tilde{S}_{i}\; .
\end{equation}

The evolution equation \eqref{parallel} for the original spin variable
$\tilde{S}_\mu$ now translates into an ordinary precession equation
for the tetrad components \eqref{constantspin}, namely
\begin{equation}\label{evolS}
\frac{\ud S_{a}}{\ud t} = \Omega_{a}^{\phantom{a}b} S_{b}\, , 
\end{equation} 
where the precession tensor is given in terms of the (covariant)
derivatives of the tetrad by
\begin{equation}\label{Omegaij}
\Omega_{a}^{\phantom{a}b} = - e_a^{\phantom{a}\mu} \frac{\uD e^b_{\phantom{b}\mu}}{\ud t} \, .
\end{equation} 
The antisymmetric character of the matrix $\Omega_{a}^{\phantom{a}b}$,
which is just made out of the usual Ricci rotation coefficients
$\omega_{\phantom{a}\nu\phantom{b}}^{a\phantom{\nu}b} = -
e^{a\mu}\nabla_\nu e^b_{\phantom{b}\mu} = -
\omega_{\phantom{b}\nu\phantom{a}}^{b\phantom{\nu}a}$, namely
$\Omega^{ab} = v^\nu
\omega_{\phantom{a}\nu\phantom{b}}^{a\phantom{\nu}b}$ where $v^\nu =
u^\nu/u^0$, guaranties that $S_a$ satisfies
an ordinary precession-type equation, \textit{i.e.}
\begin{equation}\label{prec}
	\frac{\ud \mathbf{S}}{\ud t} = \mathbf{\Omega}\times\mathbf{S} \; , 
\end{equation}
where we denote $\mathbf{S}=(S_a)$, $\mathbf{\Omega}=(\Omega_a)$ and
pose $\Omega_{a}=-\frac{1}{2}\varepsilon_{abc}\,\Omega^{bc}$. As a
consequence of \eqref{prec} the spin has a conserved Euclidean norm,
$\mathbf{S}^2=s^2$ [\textit{cf} Eq.~\eqref{orthog}].

The latter construction is completely equivalent to the one of
Ref.~\cite{Damour2008a}. Indeed the orthonormality condition
$g^{\mu\nu}\tilde{S}_{\mu}\tilde{S}_{\nu} = s^{2}$ is rewritten in
\cite{Damour2008a} as $G^{ij}\tilde{S}_{i}\tilde{S}_{j} = s^{2}$, with
an effective metric $G^{ij} = g^{ij} - 2 g^{0(i}v^{j)} + g^{00}
v^{i}v^{j}$, and the conserved norm spin variable is defined there by
$S^{i}=H^{ij}\tilde{S}_{j}$, where $H^{ij}$ is the unique symmetric
and positive definite square root of the matrix $G^{ij}$. Replacing
the metric $g^{\mu\nu}$ by the projector
$\gamma^{\mu\nu}=g^{\mu\nu}+u^\mu u^\nu$, and using
$\gamma^{\mu\nu}=\delta^{ab}e_a^{\phantom{a}\mu}e_b^{\phantom{\beta}\nu}$,
we obtain the correspondence with our previous tetrad formalism:
\begin{equation}\label{Gij}
  G^{ij} = \delta^{ab}\bigl(e_a^{\phantom{a}i} 
  - v^i e_a^{\phantom{a}0}\bigr)\bigl(e_b^{\phantom{b}j} - v^j e_b^{\phantom{b}0}\bigr)\;.
\end{equation}
Notice that the effective metric $G^{ij}$ is actually the inverse of
the spatial projector $\gamma_{ij}=g_{ij}+u_i u_j$, \textit{i.e.} we have
$G^{ij}\gamma_{jk}=\delta^i_k$. From \eqref{Gij} we get immediately
the correct factorization. Indeed, the matrix
$H_a^{\phantom{a}i}$, such that $G^{ij} =
\delta^{kl}H_k^{\phantom{k}i}H_l^{\phantom{l}j}$ in the notation of
\cite{Damour2008a}, can be written as
\begin{equation}\label{Hij}
	H_a^{\phantom{a}i}=e_a^{\phantom{a}i} - v^i e_a^{\phantom{a}0}\;.
\end{equation}
If we replace the tetrad components by \eqref{tetrad} in this
expression, we find that $H_a^{\phantom{a}i} = e_{aj}G^{ij}$. Since
the symmetric matrices $e_{ai}$ and $G^{ij}$ are, respectively, the square root
and the inverse of the matrix $\gamma_{ij}$, we obtain that the matrix
$H_a^{\phantom{a}i}$ is actually the inverse of the matrix $e_{ai}$,
namely $H_a^{\phantom{a}j}e_{bj} = \delta_{ab}$, which shows that it
is symmetric and positive definite, as required. As a result, we find that
the definition of Ref.~\cite{Damour2008a},
\begin{equation}\label{eq:DefSi}
  S_a = H_a^{\phantom{a}i} \tilde{S}_i\;.
\end{equation}
is in fact the same as our definition \eqref{constantspin}. Both
procedures explained above are therefore equivalent, but the present
approach based on a tetrad is more insightful.

\subsection{Evolution equation for the conserved spin vector}
\label{subsec:precessiongeneralframe}

From now on we shall stick to our definition \eqref{constantspin} of
the constant magnitude spin, and shall no longer make any distinction
between the spatial indices $a,b,\cdots$, which were introduced above as
tetrad indices, and the ordinary spatial indices
$i,j,k,\cdots=1,2,3$, which will be raised and lowered by the Euclidean metric. 
The 3.5PN gravitational field will be decomposed for convenience in terms of
eight metric 
potentials denoted by $V$, $V_i$, $\hat{W}_{ij}$, $\hat{R}_{i}$,
$\hat{X}$, $\hat{Z}_{ij}$, $\hat{Y}_i$ and $\hat{T}$, as introduced
in Eqs.~(3.1) of Paper~I: 
\begin{subequations}\label{metricg}
\begin{align} g_{00}
 &=  -1 + \frac{2}{c^{2}}V - \frac{2}{c^{4}} V^{2} + \frac{8}{c^{6}}
\left(\hat{X} + V_{i} V_{i} + \frac{V^{3}}{6}\right)\nonumber\\ & +
\frac{32}{c^{8}} \left(\hat{T} - \frac{1}{2} V \hat{X} + \hat{R}_{i} V_{i} -
\frac{1}{2} V V_{i} V_{i} -
\frac{V^{4}}{48}\right)+\calO\left(\frac{1}{c^{10}}
\right)\;,\\ 
g_{0i} & = - \frac{4}{c^{3}}
V_{i} - \frac{8}{c^{5}} \hat{R}_{i} - \frac{16}{c^{7}} \left(\hat{Y}_{i} +
\frac{1}{2}\hat{W}_{ij} V_{j} + \frac{1}{2} V^{2} V_{i}\right) +
\calO\left(\frac{1}{c^9}\right)\;,\\ 
g_{ij} & = \delta_{ij} \left[1 +
\frac{2}{c^{2}}V + \frac{2}{c^{4}} V^{2} + \frac{8}{c^{6}} \left(\hat{X} +
V_{k} V_{k} + \frac{V^{3}}{6}\right)\right] \nonumber\\ & +
\frac{4}{c^{4}}\hat{W}_{ij} + \frac{16}{c^{6}} \left( \hat{Z}_{ij} + \frac{1}{2} V
\hat{W}_{ij} - V_{i} V_{j} \right) + \calO\left(\frac{1}{c^8}\right) \;.
\end{align}
\end{subequations}
Each of these potentials is a retarded solution of a flat-space wave equation
sourced by components of the stress-energy tensor and appropriate lower
order potentials. We refer to Eqs.~(3.4) of Paper~I for the explicit expression
of these source terms.

Our first result is the explicit expression of the precession tensor
$\Omega^{ij}$ in terms of the metric at the 3PN relative order.  For a
single spinning particle described in the pole-dipole approximation
and moving in a fixed background, the evolution equation for the spin
is given by Eq.~\eqref{parallel}. The metric and associated
Christoffel symbols therein are replaced up to 3PN order by the
potentials $V$, $\cdots$, $\hat{Y}_i$, and the spin variable is
changed to conform with the definition \eqref{constantspin} so that we
can read off the precession tensor from Eq.~\eqref{evolS}. In practice
we use the following explicit expression in terms of the matrix
$H^{ij}$ defined by Eq.~\eqref{Hij}:
\begin{equation}\label{eq:DefAij}
  \Omega^{ij} = 
\frac{\ud H^{ik}}{\ud t} (H^{-1})^{kj} + H^{ik}K^{kl}(H^{-1})^{lj} \;,
\end{equation}
where the matrix $K^{ij}$ is a combination of velocities and
Christoffel symbols:
\begin{equation}\label{eq:DefBij}
	K^{ij} = 
c \Gamma^{j}_{\ph{j}0i} + v^{k}\Gamma^{j}_{\ph{j}ik} - 
\Bigl( \Gamma^{0}_{\ph{0}0i} + \frac{v^{k}}{c}\Gamma^{0}_{\ph{0}ik} \Bigr)v^{j} \; .
\end{equation}
We then obtain the anti-symmetric precession tensor up to 3PN order as
\begin{align}
\label{matriceAij}
	\Omega^{ij} =& \frac{1}{c^{2}} \left[ -4 \partial^{[i}V^{j]} -
3 v^{[i} \partial^{j]}V \right] + \frac{1}{c^{4}} 
\left[ -8 \partial^{[i}\hat{R}^{j]} + 8 V \partial^{[i}V^{j]} + 
4 v^{k} \partial^{[i}\hat{W}^{j]}{}_{k} -
\frac{1}{4} v^{2} v^{[i} \partial^{j]}V \right. \nn \\ 
	& \left. + 4 V^{[i} \partial^{j]}V + 2 v^{k} v^{[i} \partial^{j]}V_{k} + 
2 v^{k} v^{[i} \partial_{k}V^{j]} \vphantom{\frac{1}{2}} \right] + 
\frac{1}{c^{6}} \left[ \vphantom{\frac{1}{2}} 16 V \partial^{[i}\hat{R}^{j]} -
16 V^2 \partial^{[i}V^{j]} -16 v^{k} V_{k} \partial^{[i}V^{j]} \right. \nn \\ 
	& \left. -8 V_{k} \partial^{[i}\hat{W}^{j]}{}_{k} -
16 \partial^{[i}\hat{Y}^{j]} + 16 v^{k} \partial^{[i}\hat{Z}^{j]}{}_{k} + 
4 v^{k} v^{[i} \partial^{j]}\hat{R}_{k} + 8 \hat{R}^{[i} \partial^{j]}V - 
V v^{2} v^{[i} \partial^{j]}V \right. \nn \\ 
	& \left. -\frac{1}{8} v^{4} v^{[i} \partial^{j]}V + 
8 V V^{[i} \partial^{j]}V + v^{2} V^{[i} \partial^{j]}V -
2 v^{k} V_{k} v^{[i} \partial^{j]}V -5 v^{k} \hat{W}^{[i}{}_{k} \partial^{j]}V+ 
4 V v^{k} v^{[i} \partial^{j]}V_{k} \right. \nn \\ 
	& \left. + \frac{1}{2} v^{2} v^{k} v^{[i} \partial^{j]}V_{k} + 
8 v^{k} V^{[i} \partial^{j]}V_{k} -16 V_{k} v^{[i} \partial^{j]}V_{k} -
12 v^{[i} \partial^{j]}\hat{X} + 
4 v^{k} v^{[i} \partial_{k}\hat{R}^{j]} \right. \nn \\ 
	& \left. -4 v^{k} v^{[i} V^{j]} \partial_{k}V + 
5 v^{[i} \hat{W}^{j]k} \partial_{k}V + 4 V  v^{k} v^{[i} \partial_{k}V^{j]} + 
\frac{1}{2} v^{2} v^{k} v^{[i} \partial_{k}V^{j]} -
8 v^{k} V^{[i} \partial_{k}V^{j]} \right. \nn \\ 
	& \left. -8 V_{k} v^{[i} \partial_{k}V^{j]} + 
8 \hat{W}^{[i}{}_{k} \partial_{k}V^{j]} - 
v^{k} v^{l} v^{[i} \partial_{l}\hat{W}^{j]}{}_{k} + 
v^{k} v^{[i} \partial_{t}\hat{W}^{j]}{}_{k} \vphantom{\frac{1}{2}} \right] \; .
\end{align}
This quantity is to be evaluated at the location of the considered
particle, at which point all the potentials $V$, $V_i$, $\cdots$, that
parametrize the metric are diverging. For a system of $N$ particles,
the metric is the one generated by the stress-energy tensor of the
system of the $N$ particles itself. We consider the case $N=2$ and
compute the precession tensor \eqref{matriceAij} at the location of,
say, particle 1, the velocity appearing in \eqref{matriceAij} being
thus the ordinary velocity of particle 1.

The evaluation of \eqref{matriceAij} at 1, \textit{i.e.} at point
$\mathbf{y}_1$, is made meaningful through Hadamard's regularization
\cite{Blanchet2000}, consistently with our computation of the 3.5PN
spin-orbit acceleration in Paper~I. It must be understood in
Eq.~\eqref{eq:DefAij} that each term is regularized before taking the
products and the time derivative.  Nonetheless, although Hadamard's
regularization is ``non-distributive'' in the sense that
$(FG)_{1}\neq(F)_{1}(G)_{1}$ in general, we checked that, at the order
considered here, treating it as ``distributive'' makes no difference.
Note that the time derivative operation does not commute with the
regularization operation at 1, and we have generically for singular
functions $F$ in the class considered in
Ref.~\cite{Blanchet2000}:\footnote{This equation states that,
  formally, the Hadamard regularization commutes with the operator
  $v^{\mu}_1\partial_{\mu}$.}
\begin{equation}\label{eq:dtHadamard}
	\frac{\ud}{\ud t} (F)_{1} = (\partial_{t}F)_{1} + 
(v_{1}^{i}\partial_{i}F)_{1} \; ,
\end{equation}
where $(G)_1$ represents the value of $G$ at particle 1 position in
the sense of the Hadamard \textit{partie finie}. In order to present a
closed-form expression for $\Omega^{ij}$ in terms of the metric
potentials, we first applied the total time derivative there according
to the Leibniz rule on individual monomials composing $H^{ij}$,
applying the distributivity ansatz [\textit{i.e.}
$(FG)_{1}=(F)_{1}(G)_{1}$] for the products. We next replaced the
accelerations by their expressions in terms of the potentials. For the
time derivatives of quantities regularized at 1, we resorted to
Eq.~\eqref{eq:dtHadamard}. Finally, the partial time derivatives of
the potentials were eliminated in turn by means of the identities
(3.28) of Ref.~\cite{Blanchet2001a}, which are equivalent to the
harmonic gauge condition.
 
Since we are working at linear order in the spins, only the non-spin
parts of the metric potentials enter the computation of the matrix
$\Omega^{ij}$. Most of those contributions are the same as those
required for the 2PN equations of motion without spin.\footnote{The
  non-spin part of the acceleration has the form $a^{i} = F^{i} - \ud
  Q^{i}/\ud t $ with $Q^i = P^i - v^i$; see Eqs.~(3.5) and Eqs.~(3.7)
  in Paper~I.} There are only two genuine 3PN potentials: One of them,
$\hat{Z}_{ij}$ at Newtonian order, has the same structure as
$\hat{W}_{ij}$; The other one, $\hat{Y}_i$, which enters the term
$-16 \partial^{[i}\hat{Y}^{j]}$ in Eq.~\eqref{matriceAij}, shows a
higher order of non-linearity (in powers of $G$). Only its regularized
value can be computed, using dimensional regularization in principle,
as was done for the 3PN equations of motion without spin obtained in
\cite{Blanchet2004}. Like for the term
$\tilde{S}^{jk}(\partial_{ij}\hat{Y}_{k})_{1}$ appearing in the
equations of motion (see Section~V of Paper~I), we find that the
corrections coming from the dimensional regularization exactly cancel
out because of the antisymmetrization due to the contraction with the
spin tensor. Thus, like in Paper~I, Hadamard's regularization is
sufficient for our purpose here. The remaining 3PN metric potential,
$\hat{T}$, does not contribute.

Due to the length of the expression, we relegate to Appendix
\ref{app:Explicitspinrelation} the relation between the conserved spin
vector and the spin tensor in terms of the orbital variables derived
from Eqs.~\eqref{eq:DefSmu} and~\eqref{constantspin}. We conclude this
Section by giving the explicit expression for the precession equation
of the conserved spin 1:
\begin{equation}
\frac{\ud \mathbf{S}_1}{\ud t}=\mathbf{\Omega}_1 \times \mathbf{S}_1 \,.
\end{equation}
The vector $\mathbf{\Omega}_1$ may be expanded at 3PN order in the form:
\begin{align}
\label{Omegastruct}
\mathbf{\Omega}_1 &=
\frac{1}{c^2}\mathbf{\Omega}_1^\mathrm{1PN}
+\frac{1}{c^4}\mathbf{\Omega}_1^\mathrm{2PN}
+\frac{1}{c^6}\mathbf{\Omega}_1^\mathrm{3PN}
+ \mathcal{O}\left(\frac{1}{c^7}\right) \,.
\end{align}
Except for the spin tensor, we use the same notations for the orbital
variables as in Paper~I: $(uv)$ denotes the scalar product
$\mathbf{u}\cdot\mathbf{v}=u^i v^i$ and
$\mathbf{w}=\mathbf{u}\times\mathbf{v}$ the cross product between $\mathbf{u}$
and $\mathbf{v}$, whose components are given by $w^i=\varepsilon^{ijk}u^jv^k$.
At leading order, we have
\begin{equation}
	\mathbf{\Omega}_1^\mathrm{1PN}= \frac{G}{r_{12}^{2}} m_{2}\left[
	 \frac{3}{2} \, \mathbf{ n}_{12} \times \mathbf{v}_{1}   - 
2 \, \mathbf{ n}_{12} \times \mathbf{v}_{2} \right] \;,
\end{equation}
while the next-to-leading order correction is given by
\begin{subequations}
\begin{equation}
	\mathbf{\Omega}_1^\mathrm{2PN}= 
	\frac{G}{r_{12}^{2}}  {}^{(2)}\mathbf{\Omega}^{0,1}_1 m_{2} 
	+ \frac{G^{2}}{r_{12}^{3}} 
\left[ {}^{(2)}\mathbf{\Omega}^{0,2}_1 m_{2}^{2} + 
{}^{(2)}\mathbf{\Omega}^{1,1}_1 m_{1}m_{2} \right]  \;,
\end{equation}
where 
\allowdisplaybreaks{
\begin{align}
    {}^{(2)}\mathbf{\Omega}^{0,1}_1 &= \mathbf{n}_{12} \times \mathbf{v}_{1} \left[ -\frac{9}{4} (n_{12}v_{2})^2 + \frac{1}{8} v_{1}^{2} -  (v_{1}v_{2}) + v_{2}^{2} \right] + \mathbf{v}_{1} \times \mathbf{v}_{2} \left[ (n_{12}v_{1}) - \frac{3}{2} (n_{12}v_{2}) \right] \nonumber \\
& + \, \mathbf{n}_{12} \times \mathbf{v}_{2} \left[\vphantom{\frac{1}{2}} 3 (n_{12}v_{2})^2 + 2 (v_{1}v_{2}) - 2 v_{2}^{2} \right] \;,\nonumber \\
    {}^{(2)}\mathbf{\Omega}^{1,1}_1 &= \frac{3}{2} \, \mathbf{n}_{12} \times \mathbf{v}_{1} + \mathbf{n}_{12} \times \mathbf{v}_{2} \;, \nonumber \\
    {}^{(2)}\mathbf{\Omega}^{0,2}_1 &= -\frac{1}{2} \, \mathbf{n}_{12} \times \mathbf{v}_{1} + \frac{5}{2} \,\mathbf{n}_{12} \times \mathbf{v}_{2} \;.
\end{align}}
\end{subequations}\noindent
These results can either be directly computed from
Eq.~\eqref{matriceAij} truncated to the appropriate order or obtained
by a mere translation of the spin evolution equation given in Section
VI of Paper~I using the relation \eqref{eq:spin_conversion} of
Appendix \ref{app:Explicitspinrelation}. Using Eq.~\eqref{matriceAij},
we get the next-to-next-to-leading order correction, new with
this paper, to the precession vector:
\begin{subequations}
\begin{align}
	\mathbf{\Omega}_1^\mathrm{3PN}=& 
	\frac{G}{r_{12}^{2}} {}^{(3)}\mathbf{\Omega}^{0,1}_1 m_{2} 
	+ \frac{G^{2}}{r_{12}^{3}} 
\left[ {}^{(3)}\mathbf{\Omega}^{0,2}_1 m_{2}^{2} + 
{}^{(3)}\mathbf{\Omega}^{1,1}_1 m_{1}m_{2} \right]  \nonumber \\
	&  + \frac{G^{3}}{r_{12}^{4}} 
\left[ {}^{(3)}\mathbf{\Omega}^{0,3}_1 m_{2}^{3} + 
{}^{(3)}\mathbf{\Omega}^{1,2}_1 m_{1}m_{2}^{2}  + 
{}^{(3)}\mathbf{\Omega}^{2,1}_1 m_{1}^{2}m_{2} \right] \;,
\end{align}
where
\allowdisplaybreaks{
\begin{align}
    {}^{(3)}\mathbf{\Omega}^{0,1}_1 &= \mathbf{n}_{12} \times \mathbf{v}_{1} \left[ \frac{45}{16} (n_{12}v_{2})^4 - \frac{3}{16} (n_{12}v_{2})^2 v_{1}^{2} + \frac{1}{16} v_{1}^{4} + \frac{3}{2} (n_{12}v_{2})^2 (v_{1}v_{2}) - \frac{1}{4} v_{1}^{2} (v_{1}v_{2}) + v_{2}^{4} \right. \nonumber \\
& \left. \quad - \frac{15}{4} (n_{12}v_{2})^2 v_{2}^{2} + \frac{1}{4} v_{1}^{2} v_{2}^{2} -  (v_{1}v_{2}) v_{2}^{2}  \right] + \mathbf{n}_{12} \times \mathbf{v}_{2} \left[ -\frac{15}{4} (n_{12}v_{2})^4 + 6 (n_{12}v_{2})^2 v_{2}^{2} - 2 v_{2}^{4} \right. \nonumber \\
& \quad \left. - 3 (n_{12}v_{2})^2 (v_{1}v_{2}) + 2 (v_{1}v_{2}) v_{2}^{2} \vphantom{\frac{1}{2}} \right] + \mathbf{v}_{1} \times \mathbf{v}_{2} \left[ \frac{1}{4} (n_{12}v_{1}) v_{1}^{2} + \frac{9}{4} (n_{12}v_{2})^3 + (n_{12}v_{1}) v_{2}^{2} \right. \nonumber \\
& \quad \left. -\frac{3}{2} (n_{12}v_{1}) (n_{12}v_{2})^2  - \frac{1}{8} (n_{12}v_{2}) v_{1}^{2} - \frac{1}{2} (n_{12}v_{1}) (v_{1}v_{2}) + \frac{3}{2} (n_{12}v_{2}) (v_{1}v_{2})  - \frac{5}{2} (n_{12}v_{2}) v_{2}^{2} \right]\;, \nonumber \\
    {}^{(3)}\mathbf{\Omega}^{1,1}_1 &= \, \mathbf{n}_{12} \times \mathbf{v}_{1} \left[ -\frac{27}{2} (n_{12}v_{1})^2 + 27 (n_{12}v_{1}) (n_{12}v_{2}) - 14 (n_{12}v_{2})^2 + \frac{29}{4} v_{1}^{2} - \frac{67}{4} (v_{1}v_{2}) \right. \nonumber \\
& \quad \left. + \frac{71}{8} v_{2}^{2} \right] + \mathbf{n}_{12} \times \mathbf{v}_{2} \left[ \frac{109}{4} (n_{12}v_{1})^2 - \frac{101}{2} (n_{12}v_{1}) (n_{12}v_{2}) + \frac{73}{4} (n_{12}v_{2})^2 - \frac{37}{4} v_{1}^{2} \right. \nonumber \\
& \quad \left. + \frac{35}{2} (v_{1}v_{2}) - \frac{33}{4} v_{2}^{2} \right] + \mathbf{v}_{1} \times \mathbf{v}_{2} \left[ -\frac{71}{8} (n_{12}v_{1}) + \frac{55}{8} (n_{12}v_{2}) \right] \;,\nonumber \\
    {}^{(3)}\mathbf{\Omega}^{0,2}_1 &= \, \mathbf{n}_{12} \times \mathbf{v}_{1} \left[ \frac{1}{2} (n_{12}v_{1})^2 - \frac{3}{2} (n_{12}v_{1}) (n_{12}v_{2}) + 2 (n_{12}v_{2})^2 + \frac{3}{8} v_{1}^{2} - \frac{3}{8} (v_{1}v_{2}) - \frac{1}{2} v_{2}^{2} \right] \nonumber \\
& + \, \mathbf{n}_{12} \times \mathbf{v}_{2} \left[ -5 (n_{12}v_{2})^2 - \frac{1}{2} v_{1}^{2} - 2 (v_{1}v_{2}) + \frac{5}{2} v_{2}^{2} \right]  + \mathbf{v}_{1} \times \mathbf{v}_{2} \left[ -\frac{5}{8} (n_{12}v_{1}) - \frac{1}{2} (n_{12}v_{2}) \right] \;,\nonumber \\
    {}^{(3)}\mathbf{\Omega}^{2,1}_1 &= -\frac{69}{8} \, \mathbf{n}_{12} \times \mathbf{v}_{1} + \frac{3}{2} \, \mathbf{n}_{12} \times \mathbf{v}_{2} \;,\nonumber \\
    {}^{(3)}\mathbf{\Omega}^{1,2}_1 &= -\frac{31}{4} \, \mathbf{n}_{12} \times \mathbf{v}_{1} - \frac{17}{2} \, \mathbf{n}_{12} \times \mathbf{v}_{2} \;,\nonumber \\
    {}^{(3)}\mathbf{\Omega}^{0,3}_1 &= \frac{7}{8} \, \mathbf{n}_{12} \times \mathbf{v}_{1} - \frac{11}{2} \, \mathbf{n}_{12} \times \mathbf{v}_{2} \;.
\end{align}}
\end{subequations}

\section{Compact binaries in the center-of-mass frame}
\label{sec:CM}

\subsection{Reduction to the center-of-mass frame}
\label{subsec:CMreduction}

In this Section, we present our results in the center-of-mass (CM)
frame, defined by the nullity of the center-of-mass position
$\mathbf{G}$. We remind that the latter vector is directly linked to
the conserved integral of the motion $\mathbf{K}$ associated with the
boost invariance of the conservative part of the dynamics, through the
relation $\mathbf{K} = \mathbf{G} - \mathbf{P} t$, with $\mathbf{P}$
being the total linear momentum of the binary. We shall display
$\mathbf{G}$ in a general frame at the 3.5PN order in Appendix
\ref{appendix:CMreduction}. However, it turns out that only the 2.5PN
order is needed to reduce the results to the CM frame, since 3.5PN
corrections that would matter for the reduction of the Newtonian part
can be seen to cancel out based on symmetry arguments, as detailed at
the end of Appendix \ref{appendix:CMreduction}. When working in the CM
frame, it is convenient to use the standard mass parameters $m\equiv
m_1+m_2$, $\delta m \equiv m_1-m_2$ and $\nu\equiv m_1\,m_2/m^2$ (such
that $0<\nu\leq 1/4$), and to introduce the same combinations of spin
variables as in Refs.~\cite{Faye2006,Blanchet2006}, namely
\begin{subequations}\label{SSigma}
\begin{align}
\mathbf{S} &\equiv \mathbf{S}_1 + \mathbf{S}_2 \, , \\ \mathbf{\Sigma}
&\equiv m \Big(\frac{\mathbf{S}_2}{m_2} - \frac{\mathbf{S}_1}{m_1}
\Big)\, .
\end{align}
\end{subequations}
Notice that since we are working with different definitions of
$\mathbf{S}_1$ and $\mathbf{S}_2$, our variables $\mathbf{S}$ and
$\mathbf{\Sigma}$ also differ from those used in
Refs.~\cite{Faye2006,Blanchet2006}; however, \cite{Blanchet2006} introduced
in their Section VII conserved-norm spin variables, which turn out to be
identical to ours at 1PN order in the center-of-mass frame. Neither
 $\mathbf{S}$ nor $\mathbf{\Sigma}$ are of conserved norm and,
therefore, their evolution equations cannot reduce to precession equations;
they can straightforwardly be written in terms of the precession vectors
$\mathbf{\Omega}_1$ and $\mathbf{\Omega}_2$ as
\begin{subequations}
\begin{align}
\frac{\ud \mathbf{S}}{\ud t}&=\left(\frac{m_1}{m}\mathbf{\Omega}_1+
\frac{m_2}{m}\mathbf{\Omega}_2\right)\mathbf{\times S}+
\nu \left(\mathbf{\Omega_2-\mathbf{\Omega}_1}\right) \mathbf{\times \Sigma} \,
, \\
\frac{\ud \mathbf{\Sigma}}{\ud t}&=\left(\frac{m_2}{m}\mathbf{\Omega}_1+
\frac{m_1}{m}\mathbf{\Omega}_2\right)\mathbf{\times \Sigma}+ 
\left(\mathbf{\Omega_2-\mathbf{\Omega}_1}\right) \mathbf{\times S} \, .
\end{align}
\end{subequations}
Here, we shall simply give the CM-frame expression for $\mathbf{\Omega}_1$.

The vector $\mathbf{G}$ including spin-orbit effects at 2.5PN order
was computed in Ref.~\cite{Faye2006}. Translating it to our spin
variables and imposing $\mathbf{G}=\mathbf{0}$, we find the following
relations between the positions $\mathbf{y}_1$, $\mathbf{y}_2$ in the
CM frame and the relative position
$\mathbf{x}=\mathbf{y}_1-\mathbf{y}_2$ and velocity
$\mathbf{v}=\ud\mathbf{x}/\ud
t=\mathbf{v}_1-\mathbf{v}_2=\mathbf{v}_{12}$ (as well as spin
variables). Posing
\begin{subequations}\label{y1CM}
\begin{align}
\mathbf{y}_1 =& \frac{m_2}{m}\mathbf{x} + \mathbf{z}\, , \\
\mathbf{y}_2 =& -\frac{m_1}{m}\mathbf{x} + \mathbf{z} \, ,
\end{align}
we obtain
\begin{align}
  \mathbf{z}=&\frac{\nu}{2 c^2} \frac{\delta
    m}{m}\Bigg\{v^{2}-\frac{Gm}{r}\Bigg\}\mathbf{x} -\frac{\nu}{m c^3}
  \mathbf{ \Sigma \times v } +\frac{r \nu}{c^4}\frac{\delta
    m}{m}\Bigg\{
  \left(\frac{3}{8}-\frac{3}{2}\nu\right)v^{4}\, \mathbf{n}\nonumber\\
  &+ \frac{G m}{r}\bigg[
  \left(\left(-\frac{1}{8}+\frac{3}{4}\nu\right)(nv)^2+
    \left(\frac{19}{8}+\frac{3}{2}\nu\right)v^{2}\right)\mathbf{n} -
  \frac{7}{4}(nv)\mathbf{v}\bigg] + \frac{G^2
    m^2}{r^2}\left(\frac{7}{4}-\frac{\nu}{2}\right)
  \mathbf{n}\Bigg\}\nonumber\\
  &+\frac{\nu}{m c^5}\Bigg\{ \left(-\frac{1}{2} + 2 \nu \right)v^{2}
  \, \mathbf{ \Sigma \times v } + \frac{G m}{r}\bigg( \frac{\delta
    m}{m} \Big[ (n,S,v) \, \mathbf{n} - \frac{3}{2} (nv)\, \mathbf{ n
    \times S } - \frac{1}{2}\, \mathbf{ S \times v }
  \Big]\nonumber\\
  &\qquad\qquad+\left(- 1+ 4 \nu \right)(nv) \, \mathbf{ n \times
    \Sigma } - \left(2+\nu\right) \, \mathbf{ \Sigma \times v }\bigg)
  \Bigg\} + \mathcal{O}\left(\frac{1}{c^6}\right) \, .
\end{align}
\end{subequations}
Note that the explicit relation between the center of mass velocities
$\mathbf{v}_1$, $\mathbf{v}_2$ and the relative variables
$\mathbf{v}$, $\mathbf{x}$, as well as the two spins, can be obtained
either by taking the time derivatives of Eqs.~\eqref{y1CM} and using
the evolution equations, or equivalently by imposing that the explicit
expression of the total linear momentum $\mathbf{P}$ up to 2.5PN order
satisfies $\mathbf{P}=\mathbf{0}$, which must hold in the center of
mass frame, at that approximation level.

For each quantity of interest, we only present the (linear in) spin
part, which already leads to somewhat lengthy expressions, and refer to
the literature for the non-spin terms. However, one should keep in
mind that the spin part of a quantity in the CM frame is not the
reduction to the CM frame of the spin part of the same quantity
expressed in a general frame,
because of the crucial additional spin contributions due to the
replacement of Eqs.~\eqref{y1CM} into the non-spin part.

\subsection{Precession equation}
\label{preceq}

In the center of mass frame, the expression for the 3PN precession
vector $\mathbf{\Omega}_1$ defined by Eq.~\eqref{Omegastruct} reduces to
\begin{subequations}
\begin{align}
\label{OmegaCMstruct}
\mathbf{\Omega}_1 &=\mathbf{ n \times v } \left[
\frac{1}{c^2}\alpha_\mathrm{1PN}
+\frac{1}{c^4}\alpha_\mathrm{2PN}
+\frac{1}{c^6}\alpha_\mathrm{3PN}
+ \mathcal{O}\left(\frac{1}{c^7}\right)\right] \,,
\end{align}
\allowdisplaybreaks{
\begin{align}
\alpha_\mathrm{1PN}=&
\frac{Gm}{r^{2}}\left(\frac{3}{4} + \frac{1}{2} \nu -\frac{3}{4}\frac{\delta m}{m}\right), \\
\nonumber\\
\alpha_\mathrm{2PN}=&
\frac{Gm}{r^{2}}\left[\left(-\frac{3}{2} \nu + \frac{3}{4} \nu^2 -\frac{3}{2} \nu\frac{\delta m}{m}\right) (nv)^2+\left(\frac{1}{16} + \frac{11}{8} \nu -\frac{3}{8} \nu^2+\frac{\delta m}{m}\left(-\frac{1}{16} + \frac{1}{2} \nu\right)\right) v^{2}\right]\nonumber \\
 &+\frac{G^{2}m^{2}}{r^{3}}\left(-\frac{1}{4} -\frac{3}{8} \nu + \frac{1}{2} \nu^2+\frac{\delta m}{m}\left(\frac{1}{4} -\frac{1}{8} \nu\right)\right), \\
\nonumber\\
\alpha_\mathrm{3PN}=&
\frac{Gm}{r^{2}}\left[\left(\frac{15}{8} \nu -\frac{195}{32} \nu^2 + \frac{15}{16} \nu^3+\frac{\delta m}{m}\left(\frac{15}{8} \nu -\frac{75}{32} \nu^2\right)\right) (nv)^4\right.\nonumber \\
&\qquad\left.+\left(-3 \nu + \frac{291}{32} \nu^2 -\frac{45}{16} \nu^3+\frac{\delta m}{m}\left(-3 \nu + \frac{177}{32} \nu^2\right)\right) (nv)^2 v^{2}\right.\nonumber \\
&\qquad\left.+\left(\frac{1}{32} + \frac{19}{16} \nu -\frac{31}{8} \nu^2 + \frac{17}{16} \nu^3+\frac{\delta m}{m}\left(-\frac{1}{32} + \frac{3}{4} \nu -\frac{11}{8} \nu^2\right)\right) v^{4}\right]\nonumber \\
 &+\frac{G^{2}m^{2}}{r^{3}}\left[\left(\frac{1}{4} -\frac{525}{32} \nu -\frac{159}{16} \nu^2 + \frac{13}{4} \nu^3+\frac{\delta m}{m}\left(-\frac{1}{4} -\frac{75}{32} \nu -\frac{87}{16} \nu^2\right)\right) (nv)^2\right.\nonumber \\
&\qquad\qquad\left.+\left(\frac{3}{16} + \frac{27}{4} \nu + \frac{75}{32} \nu^2 -\frac{9}{8} \nu^3+\frac{\delta m}{m}\left(-\frac{3}{16} + \frac{9}{8} \nu + \frac{35}{32} \nu^2\right)\right) v^{2}\right]\nonumber \\
 &+\frac{G^{3}m^{3}}{r^{4}}\left(\frac{7}{16} -\frac{9}{4} \nu -\frac{9}{8} \nu^2 + \frac{1}{2} \nu^3+\frac{\delta m}{m}\left(-\frac{7}{16} -\frac{1}{8} \nu -\frac{1}{8} \nu^2\right)\right).
\end{align}}
\end{subequations}
We obtain $\mathbf{\Omega}_2$ from $\mathbf{\Omega}_1$ simply by
performing the substitution $\delta m \rightarrow - \delta m$.

\subsection{Relative acceleration}
\label{relacc}

The acceleration including spin-orbit effects up to 3.5PN order was
obtained in Paper~I. The reduction to the CM frame yields
\begin{align}\label{aCMstruct}
\frac{\ud \mathbf{v}}{\ud t} &=
\mathbf{B}_\mathrm{N}+\frac{1}{c^2}\mathbf{B}_\mathrm{1PN}+\frac{1}{c^3}
\mathop{\mathbf{B}}_{S}{}_{\!\mathrm{1.5PN}}
+\frac{1}{c^4}\left[\mathbf{B}_\mathrm{2PN}+
\mathop{\mathbf{B}}_{SS}{}_{\!\mathrm{2PN}}\right]
+\frac{1}{c^5}\left[\mathbf{B}_\mathrm{2.5PN}+
\mathop{\mathbf{B}}_{S}{}_{\!\mathrm{2.5PN}}\right] \nonumber \\
& \qquad + \frac{1}{c^6}\left[\mathbf{B}_\mathrm{3PN}+
\mathop{\mathbf{B}}_{SS}{}_{\!\mathrm{3PN}}\right] +
\frac{1}{c^7}\left[\mathbf{B}_\mathrm{3.5PN}+
  \mathop{\mathbf{B}}_{S}{}_{\!\mathrm{3.5PN}}\right]  +
\calO\left(\frac{1}{c^8}\right) \, ,
\end{align}
where we have indicated the contributions of spin-spin terms (not
considered in the present work); the non-spin terms can be found in
Ref.~\cite{Blanchet2006a}. In the rest of this paper, we use the
notation $(u,v,w)$ for the mixed product
$\varepsilon^{ijk}u^iv^jw^k$. For the spin-orbit terms, we have at
leading order the standard result \cite{KWW93,Kidder1995}
\begin{align}
  m\mathop{\mathbf{B}}_{S}{}_{\!\mathrm{1.5PN}}=& \frac{G m}{r^3}
  \left[ -6\frac{\delta m}{m} (n,\Sigma ,v) \, \mathbf{n} - 12 (n,S,v)
    \, \mathbf{n}+ 9 (nv) \, \mathbf{ n \times S }\right. \nonumber \\
  &\left.\qquad + 3\frac{\delta m}{m} (nv) \, \mathbf{ n \times \Sigma
    }+ 7 \, \mathbf{ S \times v } + 3\frac{\delta m}{m} \, \mathbf{
      \Sigma \times v }\right]\,.
\end{align}
At next-to-leading order, the results of \cite{Faye2006} may be
rewritten, with our choice of spin variables, as
\begin{subequations}
\begin{align}
m \mathop{\mathbf{B}}_{S}{}_{\!\mathrm{2.5PN}} = \frac{G m}{r^3}\left[
{}^{(2.5)}\mathbf{b}_1 + {}^{(2.5)}\mathbf{b}_2 \frac{G m}{r}\right] \, ,
\end{align} 
where
\allowdisplaybreaks{
\begin{align}
   {}^{(2.5)} \mathbf{b}_{1} &= \, \mathbf{ n \times S } \left[-\frac{45}{2} \nu (nv)^3+\left(-\frac{3}{2} + \frac{45}{2} \nu\right) (nv) v^{2}\right] + (n,\Sigma ,v) \, \mathbf{v} \frac{\delta m}{m}\left(\frac{9}{2} -6 \nu\right) (nv)\nonumber \\
& + \, \mathbf{ n \times \Sigma } \left[-15 \nu\frac{\delta m}{m} (nv)^3+\frac{\delta m}{m}\left(-\frac{3}{2} + 12 \nu\right) (nv) v^{2}\right] + (n,S,v) \, \mathbf{v} \left(\frac{21}{2} -\frac{21}{2} \nu\right) (nv) \nonumber \\
& + \, \mathbf{ S \times v } \left[\left(-\frac{3}{2} -15 \nu\right) (nv)^2+14 \nu v^{2}\right] + \, \mathbf{ \Sigma \times v } \left[\frac{\delta m}{m}\left(-\frac{3}{2} -9 \nu\right) (nv)^2+7 \nu\frac{\delta m}{m} v^{2}\right] \nonumber \\
&+ (n,\Sigma ,v) \, \mathbf{n} \left[15 \nu\frac{\delta m}{m} (nv)^2-12 \nu\frac{\delta m}{m} v^{2}\right]+ (n,S,v) \, \mathbf{n} \left[30 \nu (nv)^2-24 \nu v^{2}\right] \, ,\\
\nonumber\\ 
    {}^{(2.5)}\mathbf{b}_{2} &= (n,\Sigma ,v) \, \mathbf{n} \frac{\delta m}{m}\left(24 + \frac{37}{2} \nu\right) + (n,S,v) \, \mathbf{n} \left(44 + 33 \nu\right)+ \, \mathbf{ n \times S } \left(-28 -29 \nu\right) (nv) \nonumber \\
& + \, \mathbf{ n \times \Sigma } \frac{\delta m}{m}\left(-12 -\frac{31}{2} \nu\right) (nv) + \, \mathbf{ S \times v } \left(-24 -19 \nu\right)+ \, \mathbf{ \Sigma \times v } \frac{\delta m}{m}\left(-12 -\frac{19}{2} \nu\right) \, .\nonumber \\
\end{align}}
\end{subequations}
Finally, at next-to-next-to-leading order, we get
\begin{subequations}
\begin{align}
  m \mathop{\mathbf{B}}_{S}{}_{\!\mathrm{3.5PN}} = \frac{G
    m}{r^3}\left[ {}^{(3.5)}\mathbf{b}_1 +{}^{(3.5)}\mathbf{b}_2
    \frac{G m }{r} +{}^{(3.5)}\mathbf{b}_3 \frac{G^2
      m^2}{r^2}\right]\,,
\end{align}
where 
\allowdisplaybreaks{
\begin{align}
   {}^{(3.5)}\mathbf{b}_{1} &= (n,\Sigma ,v) \, \mathbf{n} \left[\frac{\delta m}{m}\left(-\frac{105}{4} \nu + \frac{315}{4} \nu^2\right) (nv)^4+\frac{\delta m}{m}\left(30 \nu -75 \nu^2\right) (nv)^2 v^{2}\right.\nonumber\\
    &\left.\qquad\qquad+\frac{\delta m}{m}\left(-9 \nu + 24 \nu^2\right) v^{4}\right] \nonumber \\
& + (n,S,v) \, \mathbf{n} \left[\left(-\frac{105}{2} \nu + \frac{315}{2} \nu^2\right) (nv)^4+\left(60 \nu -150 \nu^2\right) (nv)^2 v^{2}+\left(-18 \nu + 48 \nu^2\right) v^{4}\right] \nonumber \\
& + (n,\Sigma ,v) \, \mathbf{v} \left[\frac{\delta m}{m}\left(-\frac{15}{2} \nu -\frac{105}{4} \nu^2\right) (nv)^3+\frac{\delta m}{m}\left(\frac{3}{8} + \frac{15}{4} \nu + \frac{141}{8} \nu^2\right) (nv) v^{2}\right] \nonumber \\
& + (n,S,v) \, \mathbf{v} \left[\left(-\frac{15}{2} \nu -\frac{195}{4} \nu^2\right) (nv)^3+\left(\frac{3}{8} + \frac{27}{8} \nu + \frac{249}{8} \nu^2\right) (nv) v^{2}\right] \nonumber \\
& + \, \mathbf{ n \times S } \left[\left(\frac{315}{8} \nu -\frac{945}{8} \nu^2\right) (nv)^5+\left(-\frac{105}{2} \nu + \frac{585}{4} \nu^2\right) (nv)^3 v^{2}\right.\nonumber\\
    &\left.\qquad\qquad+\left(-\frac{3}{8} + \frac{165}{8} \nu -\frac{441}{8} \nu^2\right) (nv) v^{4}\right] \nonumber \\
& + \, \mathbf{ n \times \Sigma } \left[\frac{\delta m}{m}\left(\frac{105}{4} \nu -\frac{525}{8} \nu^2\right) (nv)^5+\frac{\delta m}{m}\left(-\frac{75}{2} \nu + \frac{345}{4} \nu^2\right) (nv)^3 v^{2}\right.\nonumber\\
    &\left.\qquad\qquad+\frac{\delta m}{m}\left(-\frac{3}{8} + \frac{57}{4} \nu -\frac{237}{8} \nu^2\right) (nv) v^{4}\right] \nonumber \\
& + \, \mathbf{ S \times v } \left[\left(\frac{225}{8} \nu -\frac{585}{8} \nu^2\right) (nv)^4+\left(-\frac{3}{8} -\frac{255}{8} \nu + \frac{627}{8} \nu^2\right) (nv)^2 v^{2}\right.\nonumber\\
    &\left.\qquad\qquad+\left(\frac{21}{2} \nu -28 \nu^2\right) v^{4}\right] \nonumber \\
& + \, \mathbf{ \Sigma \times v } \left[\frac{\delta m}{m}\left(15 \nu -\frac{315}{8} \nu^2\right) (nv)^4+\frac{\delta m}{m}\left(-\frac{3}{8} -\frac{69}{4} \nu + \frac{351}{8} \nu^2\right) (nv)^2 v^{2}\right.\nonumber\\
    &\left.\qquad\qquad+\frac{\delta m}{m}\left(\frac{11}{2} \nu -14 \nu^2\right) v^{4}\right] \, ,\\
    \nonumber \\\
    {}^{(3.5)}\mathbf{b}_{2} &= (n,\Sigma ,v) \, \mathbf{n} \left[\frac{\delta m}{m}\left(\frac{3147}{8} \nu + \frac{255}{4} \nu^2\right) (nv)^2+\frac{\delta m}{m}\left(-\frac{131}{8} \nu -19 \nu^2\right) v^{2}\right] \nonumber \\
& + (n,S,v) \, \mathbf{n} \left[\left(\frac{1635}{2} \nu + 117 \nu^2\right) (nv)^2+\left(-\frac{217}{4} \nu -28 \nu^2\right) v^{2}\right] \nonumber \\
& + (n,\Sigma ,v) \, \mathbf{v} \frac{\delta m}{m}\left(-\frac{381}{2} \nu -25 \nu^2\right) (nv) + (n,S,v) \, \mathbf{v} \left(-\frac{777}{2} \nu -\frac{87}{2} \nu^2\right) (nv) \nonumber \\
& + \, \mathbf{ n \times S } \left[\left(-\frac{1215}{2} \nu -105 \nu^2\right) (nv)^3+\left(\frac{1067}{4} \nu + \frac{79}{2} \nu^2\right) (nv) v^{2}\right] \nonumber \\
& + \, \mathbf{ n \times \Sigma } \left[\frac{\delta m}{m}\left(-\frac{2193}{8} \nu -\frac{279}{4} \nu^2\right) (nv)^3+\frac{\delta m}{m}\left(\frac{945}{8} \nu + 23 \nu^2\right) (nv) v^{2}\right] \nonumber \\
& + \, \mathbf{ S \times v } \left[\left(-352 \nu -\frac{123}{2} \nu^2\right) (nv)^2+\left(\frac{197}{4} \nu + 14 \nu^2\right) v^{2}\right] \nonumber \\
& + \, \mathbf{ \Sigma \times v } \left[\frac{\delta m}{m}\left(-\frac{1325}{8} \nu -\frac{147}{4} \nu^2\right) (nv)^2+\frac{\delta m}{m}\left(\frac{177}{8} \nu + 7 \nu^2\right) v^{2}\right] \, ,\\
\nonumber \\ 
    {}^{(3.5)}\mathbf{b}_{3} &= (n,\Sigma ,v) \, \mathbf{n} \frac{\delta m}{m}\left(-\frac{111}{2} -\frac{441}{4} \nu + 5 \nu^2\right) + (n,S,v) \, \mathbf{n} \left(-\frac{195}{2} -\frac{749}{4} \nu + 8 \nu^2\right) \nonumber \\
& + \, \mathbf{ n \times S } \left(\frac{121}{2} + 65 \nu -8 \nu^2\right) (nv) + \, \mathbf{ n \times \Sigma } \frac{\delta m}{m}\left(\frac{57}{2} + \frac{85}{4} \nu -6 \nu^2\right) (nv) \nonumber \\
& + \, \mathbf{ S \times v } \left(\frac{105}{2} + \frac{137}{2} \nu\right) + \, \mathbf{ \Sigma \times v } \frac{\delta m}{m}\left(\frac{57}{2} + \frac{65}{2} \nu\right).
\end{align}}
\end{subequations}

\subsection{Conserved integrals of the motion}

As already verified in Paper~I, the dynamics associated with the
next-to-next-to-leading spin orbit terms at 3.5PN order is purely
conservative (\textit{i.e.} there exists a conserved energy at that
order) and, in the harmonic coordinate system we are using, is
manifestly invariant under Poincar\'e transformations. Therefore, it
admits the usual set of Noetherian integrals of motion associated with
the Poincar\'e group (energy, linear momentum, angular momentum,
center-of-mass integral).

These integrals can be computed in a general frame from the
acceleration given in Eqs.~(6.7)--(6.12) of Paper~I (translated to our
conserved-norm spin variable) on the one hand, and the 3PN precession
equation given previously in Section
\ref{subsec:precessiongeneralframe} on the other hand, by using the
method of undetermined coefficients as was done for the energy in
Paper~I. Note that the precession equation is needed at 3PN order only
for the total angular momentum while the 2PN order is sufficient for
the other quantities.

However, a technical difficulty of the method of undetermined
coefficients is that it must deal with the fact that a given
expression does not admit a unique writing due to the existence of
dimensional identities relating the different vectors of the
problem. For this reason, we have chosen here a more direct approach
consisting in translating the expressions obtained in
Ref.~\cite{Hartung2011} from a reduced Hamiltonian formalism in
ADM-type coordinates and then \textit{verifying} that they are indeed
conserved as a consequence of our equations of motion. Note that in
the Hamiltonian formalism, the total linear momentum and the total angular
momentum are trivially expressed in terms of the canonical
variables. The translation to our variables is performed by applying
to the ADM results the contact transformation derived at the required
order in Section VII~D of Paper~I. For the angular momentum, we also
need the relation between our spin variables and the canonical spin,
1PN order beyond that given in Paper~I. The derivation of this 3PN
correspondence between both sets of spin variables is presented in
Appendix \ref{app:CompADM}.

The resulting expressions in a general frame happen to be very long,
hence we directly display them reduced to the CM frame, in which, by
definition, only the energy and the angular momentum do not
vanish. The latter quantity being still relatively lengthy, it is
relegated to Appendix \ref{app:angmom}. The energy in the CM frame
takes the form
\begin{subequations}\label{EstructCM}
\begin{align}
E &=\nu\bigg\{
e_\mathrm{N}+\frac{1}{c^2}e_\mathrm{1PN}+\frac{1}{c^3}
\mathop{e}_{S}{}_{\!\mathrm{1.5PN}}
+\frac{1}{c^4}\left[e_\mathrm{2PN}+
\mathop{e}_{SS}{}_{\!\mathrm{2PN}}\right]
+\frac{1}{c^5} \mathop{e}_{S}{}_{\!\mathrm{2.5PN}} \nonumber\\
& \qquad + \frac{1}{c^6}\left[e_\mathrm{3PN}+
\mathop{e}_{SS}{}_{\!\mathrm{3PN}}\right] + 
\frac{1}{c^7} \mathop{e}_{S}{}_{\!\mathrm{3.5PN}} + 
\mathcal{O}\left(\frac{1}{c^8}\right)\bigg\} \, ,
\end{align}
where the non-spin terms can be found for instance in
Ref.~\cite{Blanchet2006a}. For the spin-orbit contributions, we get
at leading order
\allowdisplaybreaks{
\begin{align}
\mathop{e}_{S}{}_{\!\mathrm{1.5PN}}=&
\frac{Gm}{r^{2}}\bigg\{ 
 -\frac{\delta m}{m}(n,\Sigma,v)- (n,S,v)\bigg\} \, ,
\end{align}}\noindent
which agrees with the standard results \cite{Kidder1995,Faye2006}, and at
next-to leading order 
\allowdisplaybreaks{
\begin{align}
\mathop{e}_{S}{}_{\!\mathrm{2.5PN}}=&
\frac{Gm}{r^{2}}\bigg\{ 
 (n,\Sigma,v)\frac{\delta m}{m}\left(-\frac{1}{2} + \frac{5}{2} \nu\right) v^{2}+ (n,S,v)\left[\frac{3}{2} \nu (nv)^2+\left(\frac{3}{2} + \frac{3}{2} \nu\right) v^{2}\right]\bigg\}\nonumber \\
 &+\frac{G^{2}m^{2}}{r^{3}}\nu\bigg\{ 
 -\frac{3}{2}\frac{\delta m}{m}(n,\Sigma,v)- 2(n,S,v)\bigg\}
\end{align}}\noindent
which differs from \cite{Kidder1995,Faye2006} due to the deviation in the spin
variables beyond leading order. For the next-to-next-to leading order, we
arrive at
\allowdisplaybreaks{
\begin{align}
\mathop{e}_{S}{}_{\!\mathrm{3.5PN}}=&
\frac{Gm}{r^{2}}\bigg\{ (n,\Sigma,v)\left[\frac{\delta m}{m}\left(-\frac{3}{8} + \frac{53}{8} \nu -\frac{99}{8} \nu^2\right) v^{4}
+\frac{\delta m}{m}\left(-3 \nu -\frac{3}{2} \nu^2\right) (nv)^2 v^{2}\right.\nonumber\\
    &\left.\qquad+ \frac{15}{8} \nu^2\frac{\delta m}{m} (nv)^4 \right]+ (n,S,v)\left[\left(-\frac{15}{8} \nu + \frac{45}{8} \nu^2\right) (nv)^4\right.\nonumber\\
    &\left.\qquad +\left(\frac{3}{4} \nu -\frac{51}{4} \nu^2\right) (nv)^2 v^{2}+\left(\frac{21}{8} -\frac{31}{8} \nu -\frac{55}{8} \nu^2\right) v^{4}\right]\bigg\}\nonumber \\
 &+\frac{G^{2}m^{2}}{r^{3}}\bigg\{ 
 (n,\Sigma,v)\left[\frac{\delta m}{m}\left(\frac{179}{8} \nu + \frac{3}{2} \nu^2\right) (nv)^2+\frac{\delta m}{m}\left(-2 -\frac{145}{8} \nu + \frac{41}{4} \nu^2\right) v^{2}\right]\nonumber \\ 
 &\qquad + (n,S,v)\left[\left(\frac{187}{4} \nu + 5 \nu^2\right) (nv)^2+\left(6 -\frac{143}{4} \nu + 10 \nu^2\right) v^{2}\right]\bigg\}\nonumber \\
 &+\frac{G^{3}m^{3}}{r^{4}}\bigg\{ 
 (n,\Sigma,v)\frac{\delta m}{m}\left(-\frac{1}{2} -\frac{17}{2} \nu -\frac{3}{2} \nu^2\right)+ (n,S,v)\left(-\frac{1}{2} -\frac{79}{4} \nu -2 \nu^2\right)\bigg\} \;.
\end{align}}
\end{subequations}

\subsection{Near-zone metric at the location of a particle}
\label{subsec:regularizedmetricCM}

The post-Newtonian near-zone metric up to order 3.5PN is parametrized
in terms of the elementary potentials according to
Eqs.~\eqref{metricg}. The next-to-next-to-leading spin-orbit effects
correspond to terms of order $\mathcal{O}(1/c^9)$ in $g_{00}$,
$\mathcal{O}(1/c^8)$ in $g_{0i}$ and $\mathcal{O}(1/c^7)$ in
$g_{ij}$. However, because of the complicated non-linear structure of
the source terms, analytical solutions to the appropriate order can
only be obtained for some of these potentials using the technique
described in details in Sections IIIC and D of Paper~I. As a result,
the spin-orbit contributions to the near-zone metric evaluated at an
arbitrary field point in the near zone is computable in analytic
closed form only up to $\mathcal{O}(1/c^7)$ for $g_{00}$,
$\mathcal{O}(1/c^6)$ for $g_{0i}$ and $\mathcal{O}(1/c^7)$ for
$g_{ij}$. The corresponding expressions will be given in Section
\ref{subsec:metricnearzoneCM}. By contrast, the value of all these
potentials at the location of each of the two point particles, defined
by means of the pure Hadamard-Schwartz regularization procedure, has
been calculated explicitly by using the tools presented in Section
IIID of Paper~I. This has enabled us to obtain an expression for the
regularized metric at the particle positions to the desired orders. In
addition to the regularized potentials already obtained in Paper~I, we
shall now need the spin parts of $(\hat{X})_1$ to next-to-leading
order, as well as $(\hat{T})_1$ to leading order, which we evaluate in
a similar fashion. We define
\begin{subequations}
\begin{align}\label{g00Sen1structCM}
\left(\mathop{g}_{S}{}_{00}\right)_1 &=
\frac{1}{c^5}\left(\mathop{g}_{S}{}^{\!\mathrm{1.5PN}}_{00}\right)_1
+\frac{1}{c^7}\left(\mathop{g}_{S}{}^{\!\mathrm{2.5PN}}_{00}\right)_1
+\frac{1}{c^8}\left(\mathop{g}_{S}{}^{\!\mathrm{3PN}}_{00}\right)_1
+\frac{1}{c^9}\left(\mathop{g}_{S}{}^{\!\mathrm{3.5PN}}_{00}\right)_1
+ \mathcal{O}\left(\frac{1}{c^{10}}\right)\, ,\\
\label{g0iSen1structCM}
\left(\mathop{g}_{S}{}_{0i}\right)_1 &=
\frac{1}{c^4}\left(\mathop{g}_{S}{}^{\!\mathrm{1.5PN}}_{0i}\right)_1
+\frac{1}{c^6}\left(\mathop{g}_{S}{}^{\!\mathrm{2.5PN}}_{0i}\right)_1
+\frac{1}{c^7}\left(\mathop{g}_{S}{}^{\!\mathrm{3PN}}_{0i}\right)_1
+\frac{1}{c^8}\left(\mathop{g}_{S}{}^{\!\mathrm{3.5PN}}_{0i}\right)_1
+ \mathcal{O}\left(\frac{1}{c^{9}}\right)\,,\\
\label{gijSen1structCM}
\left(\mathop{g}_{S}{}_{ij}\right)_1 &=
\frac{1}{c^5}\left(\mathop{g}_{S}{}^{\!\mathrm{2.5PN}}_{ij}\right)_1
+\frac{1}{c^7}\left(\mathop{g}_{S}{}^{\!\mathrm{3.5PN}}_{ij}\right)_1
+ \mathcal{O}\left(\frac{1}{c^{8}}\right)\, .
\end{align}
\end{subequations}
The non-spin parts can be found in Ref.~\cite{Blanchet1998}. The
  spin-orbit contributions at 3PN order, namely the terms $\sim 1/c^8$
  in ${}_{S}g_{00}$ and $\sim 1/c^7$ in ${}_{S}g_{0i}$, are actually
  pure coordinate effects. These terms can be eliminated by means of a
  gauge transformation as verified in Appendix \ref{3PNgauge}
  (extending Ref.~\cite{Blanchet2011}). After reduction to the
center-of-mass frame, we are led to the following spin parts.

\subsubsection{Metric component $g_{00}$}

\begin{subequations}
\allowdisplaybreaks{
\begin{align}
\left(\mathop{g}_{S}{}^{\!\mathrm{1.5PN}}_{00}\right)_1=&
\frac{G\nu}{r^{2}}\bigg\{ 
 (n,\Sigma,v)\left(2 +2\frac{\delta m}{m}\right)+ 4(n,S,v)\bigg\},\\
\left(\mathop{g}_{S}{}^{\!\mathrm{2.5PN}}_{00}\right)_1=&
\frac{G\nu}{r^{2}}\bigg\{ 
(n,S,v)\left[\left(-3 + 6 \nu -3\frac{\delta m}{m}\right) (nv)^2+\left(3 -8 \nu +\frac{\delta m}{m}\right) v^{2}\right]\nonumber \\
&\qquad+(n,\Sigma,v)\left[\left(-3 + 9 \nu+\frac{\delta m}{m}\left(-3 + 3 \nu\right)\right) (nv)^2\right.\nonumber\\
    &\left.\qquad\qquad\qquad+\left(2 -6 \nu+\frac{\delta m}{m}\left(2 -4 \nu\right)\right) v^{2}\right]\bigg\}\nonumber \\ 
 &+\frac{G^{2}m\nu}{r^{3}}\bigg\{ 
 (n,\Sigma,v)\left(-\frac{5}{2} + 8 \nu+\frac{\delta m}{m}\left(-\frac{17}{2} + 2 \nu\right)\right)\nonumber \\ 
 &\qquad + (n,S,v)\left(-14 + 4 \nu +\frac{\delta m}{m}\right)\bigg\},\\
\left(\mathop{g}_{S}{}^{\!\mathrm{3PN}}_{00}\right)_1=&
\frac{G^{2}m\nu}{r^{3}}\bigg\{ 
 (n,\Sigma,v)\left(2 -10\frac{\delta m}{m}\right) (nv)- 24 (nv)(n,S,v)\bigg\}, \label{g00S1CM3PN}\\
\left(\mathop{g}_{S}{}^{\!\mathrm{3.5PN}}_{00}\right)_1=&
\frac{G\nu}{r^{2}}\bigg\{ 
 (n,\Sigma,v)\left[\left(\frac{15}{4} -\frac{75}{4} \nu + \frac{75}{4} \nu^2+\frac{\delta m}{m}\left(\frac{15}{4} -\frac{45}{4} \nu + \frac{15}{4} \nu^2\right)\right) (nv)^4\right.\nonumber\\
    &\left.\qquad\qquad\qquad+\left(-6 + \frac{69}{2} \nu -48 \nu^2+\frac{\delta m}{m}\left(-6 + \frac{45}{2} \nu -15 \nu^2\right)\right) (nv)^2 v^{2}\right.\nonumber\\
    &\left.\qquad\qquad\qquad+\left(2 -13 \nu + 22 \nu^2+\frac{\delta m}{m}\left(2 -\frac{21}{2} \nu + 14 \nu^2\right)\right) v^{4}\right]\nonumber \\ 
 &\qquad + (n,S,v)\left[\left(\frac{15}{4} -15 \nu + \frac{15}{2} \nu^2+\frac{\delta m}{m}\left(\frac{15}{4} -\frac{15}{2} \nu\right)\right) (nv)^4\right.\nonumber\\
    &\left.\qquad\qquad\qquad+\left(-6 + \frac{57}{2} \nu -30 \nu^2+\frac{\delta m}{m}\left(-6 + \frac{33}{2} \nu\right)\right) (nv)^2 v^{2}\right.\nonumber\\
    &\left.\qquad\qquad\qquad+\left(\frac{11}{4} -17 \nu + 28 \nu^2+\frac{\delta m}{m}\left(\frac{5}{4} -4 \nu\right)\right) v^{4}\right]\bigg\}\nonumber \\
 &+\frac{G^{2}m\nu}{r^{3}}\bigg\{ 
 (n,\Sigma,v)\left[\left(\frac{43}{8} + \frac{39}{2} \nu -60 \nu^2+\frac{\delta m}{m}\left(\frac{119}{8} + 24 \nu -12 \nu^2\right)\right) v^{2}\right.\nonumber\\
    &\left.\qquad\qquad\qquad+\left(-\frac{145}{8} -11 \nu + 30 \nu^2+\frac{\delta m}{m}\left(-\frac{645}{8} -\frac{41}{2} \nu + 8 \nu^2\right)\right) (nv)^2\right]\nonumber \\ 
 &\qquad \qquad + (n,S,v)\left[\left(-\frac{269}{2} -20 \nu + 16 \nu^2+\frac{\delta m}{m}\left(-\frac{19}{4} - \nu\right)\right) (nv)^2\right.\nonumber\\
    &\left.\qquad\qquad\qquad\qquad+\left(\frac{59}{2} + 28 \nu -24 \nu^2+\frac{\delta m}{m}\left(\frac{25}{4} + 4 \nu\right)\right) v^{2}\right]\bigg\}\nonumber \\
 &+\frac{G^{3}m^{2}\nu}{r^{4}}\bigg\{ 
 (n,\Sigma,v)\left(-\frac{17}{2} -\frac{33}{10} \nu + 32 \nu^2+\frac{\delta m}{m}\left(-14 -\frac{31}{2} \nu + 2 \nu^2\right)\right)\nonumber \\ 
 &\qquad\qquad + (n,S,v)\left(-35 -30 \nu + 4 \nu^2+\frac{\delta m}{m}\left(-\frac{7}{2} -7 \nu\right)\right)\bigg\}.
\end{align}}
\end{subequations}

\subsubsection{Metric components $g_{0i}$}

\begin{subequations}
\allowdisplaybreaks{
\begin{align}
\left(\mathop{g}_{S}{}^{\!\mathrm{1.5PN}}_{0i}\right)_1=&
\frac{G}{r^{2}}\bigg\{\varepsilon^{ijk} n^{j} S^{k}\left(1 -\frac{\delta m}{m}\right)+ 2 \nu\varepsilon^{ijk} n^{j} \Sigma^{k}\bigg\},\\
\left(\mathop{g}_{S}{}^{\!\mathrm{2.5PN}}_{0i}\right)_1=&
\frac{G}{r^{2}}\nu\bigg\{(n,\Sigma ,v) v^{i}\left(\frac{3}{2} -3 \nu +\frac{3}{2}\frac{\delta m}{m}\right) + (n,S,v) v^{i}\left(\frac{3}{2} +\frac{3}{2}\frac{\delta m}{m}\right) \nonumber \\ 
& + \varepsilon^{ijk} n^{j} S^{k}\left[\left(-\frac{3}{2} -\frac{3}{2}\frac{\delta m}{m}\right) (nv)^2+\left(\frac{1}{2} +\frac{1}{2}\frac{\delta m}{m}\right) v^{2}\right] \nonumber \\ 
& + \varepsilon^{ijk} n^{j} \Sigma^{k}\left[\left(-\frac{3}{2} + 3 \nu -\frac{3}{2}\frac{\delta m}{m}\right) (nv)^2+\left(\frac{1}{2} - \nu +\frac{1}{2}\frac{\delta m}{m}\right) v^{2}\right] \nonumber \\ 
& + \varepsilon^{ijk} S^{j} v^{k}\left(-\frac{1}{2} -\frac{1}{2}\frac{\delta m}{m}\right) (nv)+ \varepsilon^{ijk} \Sigma^{j} v^{k}\left(-\frac{1}{2} + \nu -\frac{1}{2}\frac{\delta m}{m}\right) (nv)\bigg\}\nonumber \\
 &+\frac{G^{2}m}{r^{3}}\bigg\{\varepsilon^{ijk} n^{j} S^{k}\left(-1 +\frac{\delta m}{m}\right)+ \varepsilon^{ijk} n^{j} \Sigma^{k}\left(-2 \nu -2 \nu\frac{\delta m}{m}\right)\bigg\},\\
\left(\mathop{g}_{S}{}^{\!\mathrm{3PN}}_{0i}\right)_1=&
\frac{G^{2}m}{r^{3}}\nu\bigg\{2 (nv)\varepsilon^{ijk} n^{j} \Sigma^{k} + \frac{2}{3}\varepsilon^{ijk} \Sigma^{j} v^{k}\bigg\}, \label{g0iS1CM3PN}\\
\left(\mathop{g}_{S}{}^{\!\mathrm{3.5PN}}_{0i}\right)_1=&
\frac{G}{r^{2}}\nu\bigg\{(n,\Sigma ,v) v^{i}\left[\left(-\frac{9}{4} + 9 \nu -\frac{9}{2} \nu^2+\frac{\delta m}{m}\left(-\frac{9}{4} + \frac{9}{2} \nu\right)\right) (nv)^2\right.\nonumber\\
    &\left.\qquad\qquad+\left(\frac{13}{8} -8 \nu + \frac{37}{4} \nu^2+\frac{\delta m}{m}\left(\frac{13}{8} -\frac{19}{4} \nu\right)\right) v^{2}\right] \nonumber \\ 
& \qquad+ (n,S,v) v^{i}\left[\left(-\frac{9}{4} + \frac{27}{4} \nu+\frac{\delta m}{m}\left(-\frac{9}{4} + \frac{9}{4} \nu\right)\right) (nv)^2\right.\nonumber\\
    &\left.\qquad\qquad+\left(\frac{13}{8} -\frac{39}{8} \nu+\frac{\delta m}{m}\left(\frac{13}{8} -\frac{37}{8} \nu\right)\right) v^{2}\right] \nonumber \\ 
& \qquad+ \varepsilon^{ijk} n^{j} S^{k}\left[\left(\frac{15}{8} -\frac{45}{8} \nu+\frac{\delta m}{m}\left(\frac{15}{8} -\frac{15}{8} \nu\right)\right) (nv)^4\right.\nonumber\\
    &\left.\qquad\qquad+\left(-\frac{9}{4} + \frac{27}{4} \nu+\frac{\delta m}{m}\left(-\frac{9}{4} + \frac{21}{4} \nu\right)\right) (nv)^2 v^{2}\right.\nonumber\\
    &\left.\qquad\qquad+\left(\frac{3}{8} -\frac{9}{8} \nu+\frac{\delta m}{m}\left(\frac{3}{8} -\frac{11}{8} \nu\right)\right) v^{4}\right] \nonumber \\ 
& \qquad+ \varepsilon^{ijk} n^{j} \Sigma^{k}\left[\left(\frac{15}{8} -\frac{15}{2} \nu + \frac{15}{4} \nu^2+\frac{\delta m}{m}\left(\frac{15}{8} -\frac{15}{4} \nu\right)\right) (nv)^4\right.\nonumber\\
    &\left.\qquad\qquad+\left(-\frac{9}{4} + \frac{21}{2} \nu -\frac{21}{2} \nu^2+\frac{\delta m}{m}\left(-\frac{9}{4} + 6 \nu\right)\right) (nv)^2 v^{2}\right.\nonumber\\
    &\left.\qquad\qquad+\left(\frac{3}{8} -2 \nu + \frac{11}{4} \nu^2+\frac{\delta m}{m}\left(\frac{3}{8} -\frac{5}{4} \nu\right)\right) v^{4}\right] \nonumber \\ 
& \qquad+ \varepsilon^{ijk} S^{j} v^{k}\left[\left(\frac{3}{4} -\frac{9}{4} \nu+\frac{\delta m}{m}\left(\frac{3}{4} -\frac{3}{4} \nu\right)\right) (nv)^3\right.\nonumber\\
    &\left.\qquad\qquad+\left(-\frac{5}{8} + \frac{15}{8} \nu+\frac{\delta m}{m}\left(-\frac{5}{8} + \frac{13}{8} \nu\right)\right) (nv) v^{2}\right] \nonumber \\ 
& \qquad+ \varepsilon^{ijk} \Sigma^{j} v^{k}\left[\left(\frac{3}{4} -3 \nu + \frac{3}{2} \nu^2+\frac{\delta m}{m}\left(\frac{3}{4} -\frac{3}{2} \nu\right)\right) (nv)^3\right.\nonumber\\
    &\left.\qquad\qquad+\left(-\frac{5}{8} + 3 \nu -\frac{13}{4} \nu^2+\frac{\delta m}{m}\left(-\frac{5}{8} + \frac{7}{4} \nu\right)\right) (nv) v^{2}\right]\bigg\}\nonumber \\
 &+\frac{G^{2}m}{r^{3}}\nu\bigg\{(n,\Sigma ,v) n^{i}\left(-\frac{59}{2} + \frac{363}{4} \nu -3 \nu^2+\frac{\delta m}{m}\left(\frac{37}{2} + \frac{103}{4} \nu\right)\right) (nv) \nonumber \\ 
& \qquad+ (n,S,v) n^{i}\left(\frac{91}{2} + 50 \nu+\frac{\delta m}{m}\left(-\frac{89}{2} + \frac{3}{2} \nu\right)\right) (nv) \nonumber \\ 
& \qquad+ (n,\Sigma ,v) v^{i}\left(3 + \frac{15}{2} \nu -6 \nu^2+\frac{\delta m}{m}\left(-8 + \frac{21}{2} \nu\right)\right) \nonumber \\ 
& \qquad+ (n,S,v) v^{i}\left(-12 + 4 \nu+\frac{\delta m}{m}\left(-1 + 3 \nu\right)\right) \nonumber \\ 
& \qquad+ \varepsilon^{ijk} n^{j} S^{k}\left[\left(-\frac{131}{2} -36 \nu+\frac{\delta m}{m}\left(\frac{79}{2} -\frac{9}{2} \nu\right)\right) (nv)^2\right.\nonumber\\
    &\left.\qquad\qquad+\left(\frac{35}{2}+\frac{\delta m}{m}\left(-5 + \nu\right)\right) v^{2}\right] \nonumber \\ 
& \qquad+ \varepsilon^{ijk} n^{j} \Sigma^{k}\left[\left(11 -\frac{333}{4} \nu + 9 \nu^2+\frac{\delta m}{m}\left(-29 -\frac{97}{4} \nu\right)\right) (nv)^2\right.\nonumber\\
    &\left.\qquad\qquad+\left(-\frac{5}{4} + \frac{25}{2} \nu -2 \nu^2+\frac{\delta m}{m}\left(\frac{27}{4} + \frac{7}{2} \nu\right)\right) v^{2}\right] \nonumber \\ 
& \qquad+ \varepsilon^{ijk} S^{j} v^{k}\left(-32 -28 \nu+\frac{\delta m}{m}\left(\frac{53}{2} -\frac{3}{2} \nu\right)\right) (nv) \nonumber \\ 
& \qquad+ \varepsilon^{ijk} \Sigma^{j} v^{k}\left(\frac{43}{4} -\frac{215}{4} \nu + 3 \nu^2+\frac{\delta m}{m}\left(-\frac{57}{4} -\frac{63}{4} \nu\right)\right) (nv)\bigg\}\nonumber \\
 &+\frac{G^{3}m^{2}}{r^{4}}\bigg\{\varepsilon^{ijk} n^{j} S^{k}\left(1 + \frac{5}{4} \nu -8 \nu^2+\frac{\delta m}{m}\left(-1 -\frac{11}{4} \nu\right)\right) \nonumber \\ 
& \qquad+ \varepsilon^{ijk} n^{j} \Sigma^{k}\left(\frac{89}{20} \nu + \frac{13}{2} \nu^2+\frac{\delta m}{m}\left(-\frac{11}{20} \nu -6 \nu^2\right)\right)\bigg\}.
\end{align}}
\end{subequations}

\subsubsection{Metric components $g_{ij}$}

\begin{subequations}
\allowdisplaybreaks{
\begin{align}
\left(\mathop{g}_{S}{}^{\!\mathrm{2.5PN}}_{ij}\right)_1=&
\frac{G\nu}{r^{2}}\bigg\{v^{(i} \varepsilon^{j)kl} n^{k} \Sigma^{l} \left(2 +2\frac{\delta m}{m}\right)+ 4v^{(i}\varepsilon^{j)kl} n^{k} S^{l}\bigg\},\\
\left(\mathop{g}_{S}{}^{\!\mathrm{3.5PN}}_{ij}\right)_1=&
\frac{G\nu}{r^{2}}\bigg\{(n,\Sigma ,v) v^{i} v^{j}\left(1 -3 \nu+\frac{\delta m}{m}\left(1 - \nu\right)\right) + (n,S,v) v^{i} v^{j}\left(1 -2 \nu +\frac{\delta m}{m}\right) \nonumber \\ 
& \qquad+ v^{(i} \varepsilon^{j)kl} n^{k} \Sigma^{l} \left[\left(-3 + 9 \nu+\frac{\delta m}{m}\left(-3 + 3 \nu\right)\right) (nv)^2\right.\nonumber\\
    &\left.\qquad+\left(1 -3 \nu+\frac{\delta m}{m}\left(1 -3 \nu\right)\right) v^{2}\right] + v^{(i} \varepsilon^{j)kl} S^{k} v^{l} \left(-1 + 2 \nu -\frac{\delta m}{m}\right) (nv)\nonumber \\ 
& \qquad+ v^{(i}\varepsilon^{j)kl} n^{k} S^{l}\left[\left(-3 + 6 \nu -3\frac{\delta m}{m}\right) (nv)^2+\left(2 -6 \nu\right) v^{2}\right] \nonumber \\ 
& \qquad+ v^{(i} \varepsilon^{j)kl} \Sigma^{k} v^{l}\left(-1 + 3 \nu+\frac{\delta m}{m}\left(-1 + \nu\right)\right) (nv)\bigg\} \nonumber \\ 
 &+\frac{G^{2}m\nu}{r^{3}}\bigg\{(n,\Sigma ,v) \delta^{ij}\left(\frac{5}{2} +\frac{17}{2}\frac{\delta m}{m}\right)+ (n,\Sigma ,v) n^{i} n^{j}\left(-8 -56\frac{\delta m}{m}\right) \nonumber \\ 
& \qquad+ 17(n,S,v) \delta^{ij} + n^{(i} \varepsilon^{j)kl} n^{k} \Sigma^{l}\left(3 + 4 \nu+\frac{\delta m}{m}\left(51 + 2 \nu\right)\right) (nv)\nonumber \\ 
& \qquad - 112(n,S,v) n^{i} n^{j} + n^{(i} \varepsilon^{j)kl} n^{k} S^{l}\left(103 + 4 \nu -\frac{\delta m}{m}\right) (nv) \nonumber \\ 
& \qquad+ n^{(i} \varepsilon^{j)kl} \Sigma^{k} v^{l}\left(6 +34\frac{\delta m}{m}\right) + v^{(i}\varepsilon^{j)kl} n^{k} S^{l}\left(-23 + 4 \nu +\frac{\delta m}{m}\right)\nonumber \\ 
&\qquad + 68n^{(i} \varepsilon^{j)kl} S^{k} v^{l}  + v^{(i} \varepsilon^{j)kl} n^{k} \Sigma^{l} \left(-1 + 8 \nu+\frac{\delta m}{m}\left(-13 + 2 \nu\right)\right) \bigg\}.
\end{align}}
\end{subequations}

\subsection{Near-zone metric in the bulk}
\label{subsec:metricnearzoneCM}

Let us now turn to the spin part of the near-zone metric up to orders
$\mathcal{O}(1/c^7)$ for $g_{00}$, $\mathcal{O}(1/c^6)$ for $g_{0i}$
and $\mathcal{O}(1/c^7)$ for $g_{ij}$. As mentioned above, the wave
equations defining the potentials at play up to these orders can be
analytically integrated. We refer to Section V of \cite{Blanchet1998}
for a thorough presentation of the relevant method. The non-spin parts
of the metric up to the same orders can be found in
Ref.~\cite{Blanchet1998}.

The metric at a given arbitrary field point $\mathbf{x}$ is most
conveniently expressed using the distances to the two particles
$r_1=|\mathbf{x}-\mathbf{y_1}|$ and $r_2=|\mathbf{x}-\mathbf{y_2}|$,
as well as the unit vectors
$\mathbf{n}_1=(\mathbf{x}-\mathbf{y_1})/r_1$ and
$\mathbf{n}_2=(\mathbf{x}-\mathbf{y_2})/r_2$ pointing from the
particles to the field point. In order to avoid any confusion, let us
recall that, as in the rest of the present Section \ref{sec:CM},
$r\equiv r_{12}$ and $\mathbf{n}\equiv n_{12}$ respectively denote the
radial separation between the two bodies and the unit vector pointing
from body $2$ to body $1$.

One key ingredient in the computation of the relevant potentials is
the function $g\equiv\ln (r_1+r_2+r)$, which satisfies $\Delta
g=1/(r_1 r_2)$, where it is understood that the Laplacian operator
acts on functions of $\mathbf{x}$. In the spin part of the metric
presented below, $g$ appears within terms of the form
$(\mathop{\partial}_{1}\!{}_{ij}\mathop{\partial}_{2}\!{}_{k}g)$,
where $\mathop{\partial}_{1}\!{}_{i}$ and
$\mathop{\partial}_{2}\!{}_{i}$ denote the partial derivatives with
respect to $y_1^i$ and $y_2^i$. An explicit form for such terms is
straightforwardly obtained by expanding these derivatives but we shall
keep them factorized in an effort to reduce the length of our
expressions. Finally, we find that using the spin variables
$\mathbf{S}_1$ and $\mathbf{S_2}$ instead of their combinations
$\mathbf{\Sigma}$ and $\mathbf{S}$ leads to much more compact
expressions and, since the metric has to be symmetric under the
exchange $1 \leftrightarrow 2$, we can then further shorten our
formulae by only explicitly giving the contribution containing
$\mathbf{S}_1$. Note that in practice, the operation $1
\leftrightarrow 2$ implies the replacements $\delta m \rightarrow
-\delta m$, $\mathbf{v}\rightarrow -\mathbf{v}$ and
$\mathbf{n}\rightarrow -\mathbf{n}$ in addition to those where the
indices $1$ and $2$ explicitly appear.

\subsubsection{Metric component 00}
\allowdisplaybreaks{
\begin{align}
\mathop{g}_{S}{}_{00}=&
\frac{1}{c^5}\frac{G}{r_{1}^{2}}(n_1,S_1,v)\left(-2 +2\frac{\delta m}{m}\right)
+\frac{1}{c^7}\Bigg[\nonumber \\ 
&\frac{G}{r_{1}^{2}}(n_1,S_1,v)\left[\left(-2 + 6 \nu+\frac{\delta m}{m}\left(2 -4 \nu\right)\right) v^{2}+\left(3 -9 \nu+\frac{\delta m}{m}\left(-3 + 3 \nu\right)\right) (n_1v)^2\right]\nonumber \\
 &+\frac{G^{2}m}{r_{1}^{2}r_{2}}(n_1,S_1,v)\left(4 -4\frac{\delta m}{m}\right)+\frac{G^{2}m}{r_{1}^{3}}4\nu(n_1,S_1,v)\nonumber \\
 &+\frac{G^{2}m}{rr_{1}^{2}}\nu\bigg\{(n_1,S_1,v)\left(-8 +2\frac{\delta m}{m}\right)  - 2\frac{\delta m}{m} (nv)(n_1,n,S_1)\bigg\}\nonumber \\
 &+\frac{G^{2}m}{r^{2}r_{2}}\bigg\{(n,S_1,v)\left(2 -4 \nu+\frac{\delta m}{m}\left(-2 + 4 \nu\right)\right)+ (n_2,n,S_1)\left(\nu - \nu\frac{\delta m}{m}\right) (n_2v)\bigg\}\nonumber \\
 &+\frac{G^{2}m}{r^{2}r_{1}}\bigg\{
(n,S_1,v)\left(-3 + 2 \nu+\frac{\delta m}{m}\left(3 -4 \nu\right)\right) + (n_1,S_1,v)\left(-1 + 2 \nu +\frac{\delta m}{m}\right) (n_1n) \nonumber \\ 
&\qquad\qquad + (n_1,n,S_1)\left(2 -5 \nu+\frac{\delta m}{m}\left(-2 + \nu\right)\right) (n_1v)\bigg\}\nonumber \\
 &+\frac{G^{2}m}{r^{3}}\bigg\{(n,S_1,v)\left[\left(3 -3\frac{\delta m}{m}\right) (n_1n)+\left(-3 +3\frac{\delta m}{m}\right) (n_2n)\right] \nonumber \\ 
&\qquad\qquad + (n_1,S_1,v)\left(-\frac{5}{2} +\frac{5}{2}\frac{\delta m}{m}\right) + (n_2,S_1,v)\left(2 -2\frac{\delta m}{m}\right) \nonumber \\ 
& \qquad\qquad + (n_1,n,S_1)\left(-\frac{9}{2} +\frac{9}{2}\frac{\delta m}{m}\right) (nv) + (n_2,n,S_1)\left(3 -3\frac{\delta m}{m}\right) (nv)\bigg\}\nonumber \\ &
+G^{2}m\bigg\{16 \nu (\mathop{\partial}_{1}{}_{ja}\mathop{\partial}_{2}{}_{b}g) \varepsilon^{iab} S_1^{i} v^{j}
+ (\mathop{\partial}_{1}{}_{b}\mathop{\partial}_{2}{}_{ja}g) \varepsilon^{iab} S_1^{i} v^{j} \left(-4 + 4 \frac{\delta m}{m}\right)\nonumber \\ 
& \qquad\qquad + (\mathop{\partial}_{1}{}_{b}\mathop{\partial}_{2}{}_{aa}g) \varepsilon^{ijb} S_1^{i} v^{j} \left(4 - 8\nu - 4 \frac{\delta m}{m}\right)\bigg\}\Bigg] + (1\leftrightarrow 2) + \mathcal{O}\left(\frac{1}{c^8}\right).
\end{align}}

\subsubsection{Metric components 0i}
\allowdisplaybreaks{
\begin{align}
\mathop{g}_{S}{}_{0i}=&
\frac{2}{c^4}\frac{G}{r_{1}^{2}}\varepsilon^{ijk} n_1^{j} S_1^{k}+\frac{1}{c^6}\Bigg[\nonumber \\ 
&\frac{G}{r_{1}^{2}}\bigg\{(n_1,S_1,v) v^{i}\left(\frac{3}{2} -3 \nu -\frac{3}{2}\frac{\delta m}{m}\right)+ \varepsilon^{ijk} S_1^{j} v^{k} \left(-\frac{1}{2} + \nu +\frac{1}{2}\frac{\delta m}{m}\right) (n_1v) \nonumber \\ 
& \qquad + \varepsilon^{ijk} n_1^{j} S_1^{k}\left[\left(\frac{1}{2} - \nu -\frac{1}{2}\frac{\delta m}{m}\right) v^{2}+\left(-\frac{3}{2} + 3 \nu +\frac{3}{2}\frac{\delta m}{m}\right) (n_1v)^2\right]\bigg\}\nonumber \\
 &+\frac{G^{2}m}{r_{1}^{3}}\varepsilon^{ijk} n_1^{j} S_1^{k}\left(-1 -\frac{\delta m}{m}\right)+\frac{G^{2}m}{rr_{1}^{2}}\varepsilon^{ijk} n_1^{j} S_1^{k}\left(-2 +2\frac{\delta m}{m}\right)+\frac{G^{2}m}{r^{2}r_{2}}4\nu\varepsilon^{ijk} n^{j} S_1^{k}\nonumber \\
 &+\frac{G^{2}m}{r^{2}r_{1}}\bigg\{\varepsilon^{ijk} n^{j} S_1^{k}\left(-\frac{1}{2} -4 \nu +\frac{1}{2}\frac{\delta m}{m}\right) + \varepsilon^{ijk} n_1^{j} S_1^{k}\left(\frac{1}{2} -\frac{1}{2}\frac{\delta m}{m}\right) (n_1n)\bigg\}\nonumber \\ &
+4G^2 m (\mathop{\partial}_{1}{}_{ia}\mathop{\partial}_{2}{}_{b}g) \varepsilon^{jab} S_1^{j}\left(1 -\frac{\delta m}{m}\right)
\Bigg]+ (1\leftrightarrow 2) + \mathcal{O}\left(\frac{1}{c^7}\right).
\end{align}}

\subsubsection{Metric components ij}
\allowdisplaybreaks{
\begin{align}\label{gijSCM}
\mathop{g}_{S}{}_{ij}=&
\frac{1}{c^5}
\frac{G}{r_{1}^{2}}v^{(i} \varepsilon^{j)kl} n_1^{k} S_1^{l}\left(-2 +2\frac{\delta m}{m}\right)
+\frac{1}{c^7}\Bigg[\frac{G}{r_{1}^{2}}\bigg\{\nonumber \\ 
&
(n_1,S_1,v) v^{i} v^{j}\left(-1 + 3 \nu+\frac{\delta m}{m}\left(1 - \nu\right)\right) 
+ v^{(i} \varepsilon^{j)kl} S_1^{k} v^{l}\left(1 -3 \nu+\frac{\delta m}{m}\left(-1 + \nu\right)\right) (n_1v)\nonumber \\ 
& + v^{(i} \varepsilon^{j)kl} n_1^{k} S_1^{l}\left[\left(-1 + 3 \nu+\frac{\delta m}{m}\left(1 -3 \nu\right)\right) v^{2}+\left(3 -9 \nu+\frac{\delta m}{m}\left(-3 + 3 \nu\right)\right) (n_1v)^2\right] \bigg\}\nonumber \\
 &+\frac{G^{2}m}{r_{1}^{2}r_{2}}\bigg\{
 8 \nu(n_1,S_1,v) \delta^{ij}
 + v^{(i} \varepsilon^{j)kl} n_1^{k} S_1^{l}\left(-4 -8 \nu +4\frac{\delta m}{m}\right)
 \bigg\}+\frac{G^{2}m}{r_{1}^{3}}4 \nu v^{(i} \varepsilon^{j)kl} n_1^{k} S_1^{l}\nonumber \\
 &+\frac{G^{2}m}{rr_{1}^{2}}\nu\bigg\{
 -8(n_1,S_1,v) \delta^{ij}
 + 2\frac{\delta m}{m} (nv)n^{(i} \varepsilon^{j)kl} n_1^{k} S_1^{l} 
 + 2\frac{\delta m}{m}v^{(i} \varepsilon^{j)kl} n_1^{k} S_1^{l}\bigg\}\nonumber \\
 &+\frac{G^{2}m}{r^{2}r_{2}}\bigg\{
 (n_2,n,S_1) \delta^{ij}\left(\nu - \nu\frac{\delta m}{m}\right) (n_2v)
 + (n,S_1,v) \delta^{ij}\left(2 -2\frac{\delta m}{m}\right)\nonumber \\
 &+ v^{(i} \varepsilon^{j)kl} n^{k} S_1^{l}\left(-4 \nu +4 \nu\frac{\delta m}{m}\right)
 \bigg\}+\frac{G^{2}m}{r^{2}r_{1}}\bigg\{
 (n_1,n,S_1) \delta^{ij}\left(- \nu +\nu\frac{\delta m}{m}\right) (n_1v)\nonumber \\ 
&+ 8 \nu(n,S_1,v) \delta^{ij} + n^{(i} \varepsilon^{j)kl} n_1^{k} S_1^{l}\left(-2 + 4 \nu +2\frac{\delta m}{m}\right) (n_1v)
+ n^{(i} \varepsilon^{j)kl} S_1^{k} v^{l}\left(-4 +4\frac{\delta m}{m}\right) \nonumber \\ 
& + v^{(i} \varepsilon^{j)kl} n_1^{k} S_1^{l}\left(-1 + 2 \nu +\frac{\delta m}{m}\right) (n_1n) 
+ v^{(i} \varepsilon^{j)kl} n^{k} S_1^{l}\left(1 -6 \nu+\frac{\delta m}{m}\left(-1 -4 \nu\right)\right)\bigg\}\nonumber \\
 &+\frac{G^{2}m}{r^{3}}\bigg\{
 (n_1,n,S_1) \delta^{ij}\left(-\frac{3}{2} +\frac{3}{2}\frac{\delta m}{m}\right) (nv)
 + (n_2,n,S_1) \delta^{ij}\left(3 -3\frac{\delta m}{m}\right) (nv) \nonumber \\ 
& + (n_1,S_1,v) \delta^{ij}\left(-\frac{3}{2} +\frac{3}{2}\frac{\delta m}{m}\right) 
+ (n_2,S_1,v) \delta^{ij}\left(2 -2\frac{\delta m}{m}\right) \nonumber \\ 
&+ (n,S_1,v) \delta^{ij}\left[\left(3 -3\frac{\delta m}{m}\right) (n_1n)+\left(-3 +3\frac{\delta m}{m}\right) (n_2n)\right] \nonumber \\ 
&+ n^{(i} \varepsilon^{j)kl} n_1^{k} S_1^{l}\left(3 -3\frac{\delta m}{m}\right) (nv) 
+ v^{(i} \varepsilon^{j)kl} n_1^{k} S_1^{l}\left(-1 +\frac{\delta m}{m}\right)
\bigg\}\nonumber \\ 
&+G^{2}m\bigg\{
4 \delta^{ij} (\mathop{\partial}_{1}{}_{b}\mathop{\partial}_{2}{}_{la}g) \varepsilon^{kab} S_1^{k} v^{l} \left(-1 + \frac{\delta m}{m}\right)
-16 \nu \delta^{ij} (\mathop{\partial}_{1}{}_{ab}\mathop{\partial}_{2}{}_{a}g) \varepsilon^{klb} S_1^{k} v^{l}\nonumber \\ 
&+ \delta^{ij} (\mathop{\partial}_{1}{}_{b}\mathop{\partial}_{2}{}_{aa}g) \varepsilon^{klb} S_1^{k} v^{l} \left(4 - 8\nu - 4 \frac{\delta m}{m}\right)
+ 8 (\mathop{\partial}_{1}{}_{ka}\mathop{\partial}_{2}{}_{(j}g) \varepsilon^{i)ba} S_1^{b} v^{k} \left(1 - \frac{\delta m}{m}\right)\nonumber \\ 
&+ 8 (\mathop{\partial}_{1}{}_{(ia}\mathop{\partial}_{2}{}_{j)}g) \varepsilon^{bka} S_1^{b} v^{k} \left(1 - \frac{\delta m}{m}\right)
+ 16 \nu (\mathop{\partial}_{1}{}_{(ja}\mathop{\partial}_{2}{}_{b}g) \varepsilon^{kab} S_1^{k} v^{i)} \nonumber \\ 
&- 16 \nu (\mathop{\partial}_{1}{}_{ab}\mathop{\partial}_{2}{}_{a}g) \varepsilon^{(jkb} S_1^{k} v^{i)} 
\bigg\}
\Bigg]+ (1\leftrightarrow 2) + \mathcal{O}\left(\frac{1}{c^8}\right).
\end{align}}

\section{Compact binaries in quasi-circular orbits}
\label{sec:CO}

\subsection{Reduction to circular orbits}
\label{subsec:COreductiondefs}

We now present our results in the case where the orbit is nearly
circular (if one neglects the radiation-reaction effects),
\textit{i.e.} has a constant radius and is planar apart from small
perturbations induced by the spins that cause the orbital plane to
slowly precess. Most compact binaries will be in quasi-circular orbits
when they are observed by the gravitational wave detectors. Assuming
this sort of motion drastically simplifies the long expressions of the
previous Section, which become immediately much shorter. We may thus
include also, for the sake of completeness and convenience, all
non-spinning terms up to 3PN order. The results that we display are
then complete up to 3.5PN order and linear order in the spins
(remember that we neglect spin-spin contributions).\footnote{Note that
  in order to reduce the results of the previous Section to circular
  orbits, the contributions from the non-spinning terms up to 2PN
  order, which are not presented there, must be taken into account.}

When comparing with the various expressions in the literature for the
lower order spin-orbit corrections, we have to keep in mind that
different definitions of the spin variables yield different numerical
coefficients. In particular, as Ref.~\cite{Faye2006} and Section VI of
Ref.~\cite{Blanchet2006} did not make use of conserved-norm spins, the
results there differ from the ones that we find here at
next-to-leading order. By contrast, our energy, angular momentum and
precession vector agree with those given in Section VII of
\cite{Blanchet2006}, where conserved-norm spin variables are
introduced, because these spin variables, although constructed
differently, turn out to agree with ours when evaluated in the CM
frame.

For the kinematics of quasi-circular orbits, we use the same
conventions as Ref.~\cite{Faye2006} which can be summarized as
follows. We introduce an orthonormal triad
$\{\mathbf{n},\bm{\lambda},\bm{\ell}\}$, such that
$\mathbf{n}=\mathbf{x}/r$,
$\bm{\ell}=\mathbf{L}_\mathrm{N}/\vert\mathbf{L}_\mathrm{N}\vert$
where $\mathbf{L}_\mathrm{N}\equiv\mu\,\mathbf{x}\times\mathbf{v}$
denotes the Newtonian angular momentum, and
$\bm{\lambda}=\bm{\ell}\times\mathbf{n}$. All the relevant CM
quantities have been defined in Section \ref{subsec:CMreduction}. We
project out the spins on the latter orthonormal basis, as $\mathbf{S}=
S_n \mathbf{n} + S_\lambda \bm{\lambda} + S_\ell \bm{\ell}$ and
similarly for $\mathbf{\Sigma}$. By definition of quasi-circular
orbits, the orbital separation decreases as $\dot{r}=\calO(1/c^5)$,
entirely due to the gravitational radiation reaction.  The spins
contribute to this radiation reaction only at 1.5PN order beyond the
dominant term $\calO(1/c^5)$, \textit{i.e.} at 4PN order
\cite{Will05}, which is neglected here. We define the orbital
frequency $\omega$ and the precession frequency $\varpi$ by the
relations $\omega = \bm{\lambda}.\dot{\bm{n}}$ and $\varpi =
\bm{\ell}.\dot{\bm{\lambda}}$, so that the triad evolution equations
read $\dot{\bm{n}} = \omega\bm{\lambda}$, $\dot{\bm{\lambda}} =
-\omega\bm{n} + \varpi \bm{\ell}$ and $\dot{\bm{\ell}} = - \varpi
\bm{\lambda}$. Then, posing $\bm{\varpi}=\varpi\bm{n}$, we have
  $\dot{\bm{\ell}} = \bm{\varpi}\times\bm{\ell}$ which is the natural
  definition for the precession vector $\bm{\varpi}$ of the orbital
  plane.\footnote{We have $\varpi=-\omega_\text{prec}$ in the
    notation of Ref.~\cite{Blanchet2011}.}  The acceleration may be
then written in our moving basis as
\begin{equation}\label{acirc}
  \frac{\ud \mathbf{v}}{\ud t} = - \omega^2 r\,\mathbf{n} + (r\dot{\omega} +
  2\dot{r}\omega)\bm{\lambda} + a_\ell\,\bm{\ell} 
  + \mathcal{O}\left(\frac{1}{c^8}\right) \, ,
\end{equation}
in which we have set $a_{\ell} \equiv r \omega \varpi$ and neglected a
term $\ddot{r}=\mathcal{O}(1/c^{8})$. The spin contributions in the
term proportional to $\bm{\lambda}$ are, by the same argument as
before, of negligible 4PN order since the spin terms in $\omega$ come
with a factor $1/c^3$ and can thus be regarded as constant at the
considered order, see Eq.~\eqref{omega2ofgamma} below and the
discussion thereafter. As a result, the spins can contribute to the
acceleration only through $\omega$ and the quantity $a_\ell$ at 3.5PN
order.

\subsection{Equations of motion and precession vector}

In the case of circular orbits, the relativistic generalization of
Kepler's law relating the orbital frequency $\omega$ to the body
separation $r$, is most conveniently expressed in terms of the
post-Newtonian parameter $\gamma\equiv G m/(r
c^2)=\mathcal{O}(1/c^2)$. It is derived by reducing the CM relative
acceleration given in Section \ref{relacc}. We find
\begin{align}
\label{omega2ofgamma}
\omega^2=\frac{G m}{r^3}\Bigg\{&
1
+\gamma \left(-3+\nu\right)
+\gamma^2 \left(6 + \frac{41}{4} \nu + \nu^2\right)\nonumber\\
&+\gamma^3 \left( 
-10+\left[-\frac{75707}{840}+\frac{41}{64}\pi^2+22\ln\left(\frac{r}{r'_0}\right)\right]\nu + \frac{19}{2} \nu^2 + \nu^3\right) \nonumber\\
&+\frac{\gamma^{3/2}}{G m^2} \left[-5S_\ell-3\frac{\delta m}{m}\Sigma _\ell\right]
+\frac{\gamma^{5/2}}{G m^2} \left[\left(\frac{45}{2} -\frac{27}{2} \nu\right)S_\ell+\frac{\delta m}{m}\left(\frac{27}{2} -\frac{13}{2} \nu\right)\Sigma _\ell\right]\nonumber\\
&+\frac{\gamma^{7/2}}{G m^2} \left[\left(-\frac{495}{8} -\frac{561}{8} \nu -\frac{51}{8} \nu^2\right)S_\ell+\frac{\delta m}{m}\left(-\frac{297}{8} -\frac{341}{8} \nu -\frac{21}{8} \nu^2\right)\Sigma _\ell\right]\nonumber\\
&+\mathcal{O}\left(\frac{1}{c^8}\right)
\Bigg\}.
\end{align}
This expression is clearly gauge dependent since it involves the
separation $r$ defined in harmonic coordinates. Notably, the presence
of a gauge constant $r'_0$ in the non-spin part at the 3PN order
attests the occurrence of a pure gauge effect which disappears from
the observables of the problem (see \textit{e.g.}
\cite{Blanchet2006a}). As we stated before, the spin part of $\omega$
is constant at linear order in spin and up to 3.5PN order, for $r$ is
constant at the 2PN approximation and so are the spin components
$S_\ell^A$ (with $A=1,2$): indeed, the precession vectors
$\bm{\Omega}_{A}$ are proportional to $\bm{\ell}$, see
Eq.~\eqref{precessionCO}, hence $\ud S_\ell^A/\ud t = -\varpi
S_\lambda^A$, the right-hand side of the latter relation being
quadratic in spin since $\varpi$ is proportional to the spin,
  see Eq.~\eqref{alCO} below.

When computing physical quantities, it is useful to eliminate the
coordinate distance $r$ (or equivalently $\gamma$) by reexpressing it
in terms of $\omega$ or, more conveniently, of the post-Newtonian
frequency-related parameter $x\equiv (G m \omega/c^3)^{2/3}$.
Inverting Eq.~\eqref{omega2ofgamma}, we obtain
\begin{align}
\gamma=x\Bigg\{ &
1
+x \left(1 -\frac{1}{3} \nu\right)
+x^2 \left(1 -\frac{65}{12} \nu\right)\nonumber\\
&+x^3 \left(
1+\left[-\frac{2203}{2520}-\frac{41}{192}\pi^2-\frac{22}{3}\ln\left(\frac{r}{r'_0}\right)\right]\nu + \frac{229}{36} \nu^2 + \frac{1}{81}\nu^3\right) \nonumber\\
&+\frac{x^{3/2}}{G m^2}\left[\frac{5}{3}S_\ell+\frac{\delta m}{m}\Sigma _\ell\right]
+\frac{x^{5/2}}{G m^2} \left[\left(\frac{10}{3} + \frac{8}{9} \nu\right)S_\ell+2\frac{\delta m}{m}\Sigma _\ell\right]\nonumber\\
&+\frac{x^{7/2}}{G m^2} \left[\left(5 -\frac{127}{12} \nu -6 \nu^2\right)S_\ell+\frac{\delta m}{m}\left(3 -\frac{61}{6} \nu -\frac{8}{3} \nu^2\right)\Sigma _\ell\right]\nonumber\\
&+\mathcal{O}\left(\frac{1}{c^8}\right)
\Bigg\}.
\end{align}
For the component of the acceleration along $\boldsymbol\ell$, which vanishes
in the absence of spins, after inserting the explicit value of
$\mathbf{v} = \dot{r} \mathbf{n}+ r \omega \bm{\lambda} $ and eliminating $r$
in terms of $x$ by means of the above formula, we find (recall that $a_{\ell}
= r\omega\varpi$)
\begin{align}
\label{alCO}
a_\ell=\frac{c^4 x^{7/2}}{G^2 m^3}\Bigg\{&
\left[7S_n+3\frac{\delta m}{m}\Sigma _n\right]
+x \left[\left(-10 -\frac{29}{3} \nu\right)S_n+\frac{\delta m}{m}\left(-6 -\frac{9}{2} \nu\right)\Sigma _n\right]\nonumber\\
& +x^2 \left[\left(\frac{3}{2} + \frac{59}{4} \nu + \frac{52}{9} \nu^2\right)S_n+\frac{\delta m}{m}\left(\frac{3}{2} + \frac{73}{8} \nu + \frac{17}{6} \nu^2\right)\Sigma _n\right]\Bigg\}+\mathcal{O}\left(\frac{1}{c^8}\right).
\end{align}
The last quantity that we need in order to complete the evolution
equations for circular orbits is the precession vector
$\mathbf{\Omega}_1$. From Section \ref{preceq}, we readily get
\begin{align}
\label{precessionCO}
\mathbf{\Omega}_1=\frac{c^3 x^{5/2}}{Gm}\boldsymbol\ell \Bigg\{ &
\left(\frac{3}{4} + \frac{1}{2} \nu -\frac{3}{4}\frac{\delta m}{m}\right)
+x \left[\frac{9}{16} + \frac{5}{4} \nu -\frac{1}{24} \nu^2+\frac{\delta m}{m}\left(-\frac{9}{16} + \frac{5}{8} \nu\right)\right]\nonumber\\
&+x^2 \left[\frac{27}{32} + \frac{3}{16} \nu -\frac{105}{32} \nu^2 -\frac{1}{48} \nu^3+\frac{\delta m}{m}\left(-\frac{27}{32} + \frac{39}{8} \nu -\frac{5}{32} \nu^2\right)\right]\nonumber\\
&+\mathcal{O}\left(\frac{1}{c^6}\right)
\Bigg\}.
\end{align}

\subsection{Conserved integrals of the motion}

Let us now provide the expressions of the energy $E$ and of the
orbital angular momentum $\mathbf{L}$ reduced to the case of circular
motion. Applying to the CM expression \eqref{EstructCM} the same
reduction procedure as for the computation of $a_\ell$ in the previous
Subsection yields
\begin{align}
\label{Eofx}
E=-\frac{\mu c^2 x}{2}\Bigg\{ &
1
+x \left(-\frac{3}{4} -\frac{1}{12} \nu\right)
+x^2 \left(-\frac{27}{8} + \frac{19}{8} \nu -\frac{1}{24} \nu^2\right)\nonumber\\
&+x^3 \left( -\frac{675}{64}+\left[\frac{34445}{576}-\frac{205}{96}\pi^2\right]\nu-\frac{155}{96}\nu^2-\frac{35}{5184}\nu^3\right) \nonumber\\
&+\frac{x^{3/2}}{G m^2}\left[\frac{14}{3}S_\ell+2\frac{\delta m}{m}\Sigma _\ell\right]
+\frac{x^{5/2}}{G m^2} \left[\left(11 -\frac{61}{9} \nu\right)S_\ell+\frac{\delta m}{m}\left(3 -\frac{10}{3} \nu\right)\Sigma _\ell\right]\nonumber\\
&+\frac{x^{7/2}}{G m^2} \left[\left(\frac{135}{4} -\frac{367}{4} \nu + \frac{29}{12} \nu^2\right)S_\ell+\frac{\delta m}{m}\left(\frac{27}{4} -39 \nu + \frac{5}{4} \nu^2\right)\Sigma _\ell\right]\nonumber\\
&+\mathcal{O}\left(\frac{1}{c^8}\right)
\Bigg\} \;,
\end{align}
where the non-spin terms are taken from
Ref.~\cite{Blanchet2006a}. Notice that since $E$ is a physical
observable, the gauge constant $r'_0$ has canceled out as
expected. For the total angular momentum, we obtain from the CM
expression provided in Appendix \ref{app:angmom}:
\begin{align}
\mathbf{L}=\frac{G m^2 }{c\,  x^{1/2}}\nu \Bigg\{ &
\boldsymbol\ell\left[
1
+x \left(\frac{3}{2} + \frac{1}{6} \nu\right)
+x^2 \left(\frac{27}{8} -\frac{19}{8} \nu + \frac{1}{24} \nu^2\right)\right.\nonumber\\
&\qquad\left.+x^3 \left(\frac{135}{16} +\left[- \frac{6889}{144} + \frac{41}{24} \pi^2\right] \nu +\frac{31}{24} \nu^2 + \frac{7}{1296} \nu^3 \right)\right] \nonumber\\
&+\frac{x^{3/2}}{G m^2}\Bigg(
\boldsymbol\ell\left[-\frac{35}{6}S_\ell-\frac{5}{2}\frac{\delta m}{m}\Sigma _\ell\right]
+\boldsymbol\lambda\left[-3S_{\lambda }-\frac{\delta m}{m}\Sigma _{\lambda }\right]
+\mathbf{n}\left[\frac{1}{2}S_n+\frac{1}{2}\frac{\delta m}{m}\Sigma _n\right]
\Bigg)\nonumber\\
&+\frac{x^{5/2}}{G m^2}\Bigg(
\boldsymbol\ell\left[\left(-\frac{77}{8} + \frac{427}{72} \nu\right)S_\ell+\frac{\delta m}{m}\left(-\frac{21}{8} + \frac{35}{12} \nu\right)\Sigma _\ell\right]\nonumber\\
&\qquad\qquad+\boldsymbol\lambda\left[\left(-\frac{7}{2} + 3 \nu\right)S_{\lambda }+\frac{\delta m}{m}\left(-\frac{1}{2} + \frac{4}{3} \nu\right)\Sigma _{\lambda }\right]\nonumber\\
&\qquad\qquad+\mathbf{n}\left[\left(\frac{11}{8} -\frac{19}{24} \nu\right)S_n+\frac{\delta m}{m}\left(\frac{11}{8} -\frac{5}{12} \nu\right)\Sigma _n\right]
\Bigg)\nonumber\\
&+\frac{x^{7/2}}{G m^2}\Bigg(
\boldsymbol\ell\left[\left(-\frac{405}{16} + \frac{1101}{16} \nu -\frac{29}{16} \nu^2\right)S_\ell+\frac{\delta m}{m}\left(-\frac{81}{16} + \frac{117}{4} \nu -\frac{15}{16} \nu^2\right)\Sigma _\ell\right]\nonumber\\
&\qquad\qquad
+\boldsymbol\lambda\left[\left(-\frac{29}{4} + \frac{1}{12} \nu -\frac{4}{3} \nu^2\right)S_{\lambda }+\frac{\delta m}{m}\left(-\frac{1}{2} -\frac{79}{24} \nu -\frac{2}{3} \nu^2\right)\Sigma _{\lambda }\right]\nonumber\\
&\qquad\qquad
+\mathbf{n}\left[\left(\frac{61}{16} -\frac{1331}{48} \nu + \frac{11}{48} \nu^2\right)S_n+\frac{\delta m}{m}\left(\frac{61}{16} -\frac{367}{24} \nu + \frac{5}{48} \nu^2\right)\Sigma _n\right]
\Bigg)\nonumber\\
&+\mathcal{O}\left(\frac{1}{c^8}\right)
\Bigg\} \;.
\end{align}

\noindent
The non-spin parts can be found for instance in Ref.~\cite{LeTiec2012}. At the
next-to-leading order in spin-orbit effects, for both $E$ and $\mathbf{L}$, we
recover the results of Section VII of Ref.~\cite{Blanchet2006}, which are
given in terms of the conserved-norm spin variables defined there.

\subsection{Metric regularized at a particle location}

Finally, we give the expression of the spin part of the near-zone metric
regularized at the location of body $1$ for circular orbits:
\begin{subequations}
\begin{align}
\left(\mathop{g}_{S}{}_{00}\right)_1=
&\frac{x^{5/2}}{G m^2}\left[\left(\frac{5}{3} -4 \nu -\frac{5}{3}\frac{\delta m}{m}\right)S_\ell+\left(-1 + 2 \nu+\frac{\delta m}{m}\left(1 -2 \nu\right)\right)\Sigma _\ell\right]\nonumber\\
&+\frac{x^{7/2}}{G m^2}\nu\left[\left(\frac{11}{9} + \frac{16}{3} \nu -\frac{56}{9}\frac{\delta m}{m}\right)S_\ell+\left(-\frac{7}{2} + \frac{20}{3} \nu+\frac{\delta m}{m}\left(\frac{1}{2} + \frac{8}{3} \nu\right)\right)\Sigma _\ell\right]\nonumber\\
&+\frac{x^{9/2}}{G m^2}\nu\left[\left(\frac{9}{4} + \frac{101}{3} \nu -\frac{8}{3} \nu^2+\frac{\delta m}{m}\left(-\frac{45}{2} + \frac{118}{3} \nu\right)\right)S_\ell\right.\nonumber\\
&\qquad\qquad+\left.\left(-\frac{13}{8} + \frac{1009}{30} \nu -64 \nu^2+\frac{\delta m}{m}\left(-\frac{41}{8} + \frac{79}{6} \nu -\frac{4}{3} \nu^2\right)\right)\Sigma _\ell\right]\nonumber\\
&+\mathcal{O}\left(\frac{1}{c^{10}}\right) \;,
\end{align}
\begin{align}
\left(\mathop{g}_{S}{}_{0i}\right)_1=
&\frac{x^2}{G m^2}\Bigg(
\ell^i \left[\left(1 -\frac{\delta m}{m}\right)S_{\lambda }+2 \nu\Sigma _{\lambda }\right]
+\lambda^i\left[\left(-1 +\frac{\delta m}{m}\right)S_\ell-2 \nu\Sigma _\ell\right]
\Bigg)\nonumber\\
&+\frac{x^3}{G m^2}\Bigg(
\ell^i \left[\left(1 -\frac{1}{6} \nu+\frac{\delta m}{m}\left(-1 + \frac{7}{6} \nu\right)\right)S_{\lambda }+\left(\frac{5}{2} \nu -\frac{7}{3} \nu^2 -\frac{3}{2} \nu\frac{\delta m}{m}\right)\Sigma _{\lambda }\right]\nonumber\\
&\qquad\qquad
+\lambda^i\left[\left(-1 -\frac{4}{3} \nu+\frac{\delta m}{m}\left(1 -\frac{8}{3} \nu\right)\right)S_\ell+\left(-4 \nu + \frac{16}{3} \nu^2\right)\Sigma _\ell\right]\Bigg)\nonumber\\
&+\frac{x^{7/2}}{G m^2}\nu \Bigg(-\frac{2}{3} \Sigma _\ell  n^{i} + \frac{2}{3}   \Sigma _n \ell^{i}\Bigg)\nonumber\\
&+\frac{x^4}{G m^2}\Bigg(
\ell^i \left[\left(1 + \frac{69}{8} \nu -\frac{649}{72} \nu^2+\frac{\delta m}{m}\left(-1 + \frac{25}{8} \nu -\frac{35}{72} \nu^2\right)\right)S_{\lambda }\right.\nonumber\\
&\qquad\qquad\qquad+\left.\left(\frac{143}{40} \nu -4 \nu^2 + \frac{35}{36} \nu^3+\frac{\delta m}{m}\left(\frac{23}{40} \nu -\frac{7}{4} \nu^2\right)\right)\Sigma _{\lambda }\right]\nonumber\\
&\qquad\qquad
+\lambda^i\left[\left(-1 -\frac{23}{4} \nu + \frac{389}{9} \nu^2+\frac{\delta m}{m}\left(1 -\frac{25}{4} \nu + \frac{19}{9} \nu^2\right)\right)S_\ell\right.\nonumber\\
&\qquad\qquad\qquad+\left.\left(-\frac{97}{10} \nu + \frac{21}{2} \nu^2 -\frac{38}{9} \nu^3+\frac{\delta m}{m}\left(\frac{13}{10} \nu + 16 \nu^2\right)\right)\Sigma _\ell\right]\Bigg)\nonumber\\
&+\mathcal{O}\left(\frac{1}{c^9}\right) \;,
\end{align}
\begin{align}
\left(\mathop{g}_{S}{}_{ij}\right)_1=&
\frac{x^{5/2}}{G m^2}\Bigg(
\delta^{ij} \left[\left(\frac{5}{3} -\frac{5}{3}\frac{\delta m}{m}\right)S_\ell+\left(-1 + 4 \nu +\frac{\delta m}{m}\right)\Sigma _\ell\right]\nonumber\\
&
+\ell^{(i}\lambda^{j)}\nu\left[4S_{\lambda }+\left(2 +2\frac{\delta m}{m}\right)\Sigma _{\lambda }\right]
+\lambda^{i}\lambda^{j}\nu\left(-2 \Sigma _\ell -4 S_\ell -2 \Sigma _\ell\frac{\delta m}{m}\right)\Bigg)\nonumber\\
&+\frac{x^{7/2}}{G m^2}\Bigg(
n^i n^j \left[\left(\frac{5}{3} + 14 \nu -\frac{5}{3}\frac{\delta m}{m}\right)S_\ell+\left(-1 + 6 \nu+\frac{\delta m}{m}\left(1 + 4 \nu\right)\right)\Sigma _\ell\right]\nonumber\\
&
+\delta^{ij} \left[\left(5 -\frac{145}{9} \nu+\frac{\delta m}{m}\left(-5 -\frac{8}{9} \nu\right)\right)S_\ell+\left(-3 + \frac{19}{2} \nu+\frac{\delta m}{m}\left(3 -\frac{17}{2} \nu\right)\right)\Sigma _\ell\right]\nonumber\\
&
+\ell^{(i}\lambda^{j)}\nu\left[\left(-17 -\frac{10}{3} \nu +\frac{\delta m}{m}\right)S_{\lambda }+\left(2 + \frac{13}{3} \nu+\frac{\delta m}{m}\left(-10 -\frac{5}{3} \nu\right)\right)\Sigma _{\lambda }\right]\nonumber\\
&
+\lambda^{i}\lambda^{j}\nu\left[\left(\frac{38}{3} + \frac{16}{3} \nu -\frac{16}{3}\frac{\delta m}{m}\right)S_\ell+\left(-5 + \frac{20}{3} \nu+\frac{\delta m}{m}\left(7 + \frac{8}{3} \nu\right)\right)\Sigma _\ell\right]\nonumber\\
&
+\ell^{(i} n^{j)} \nu\left[68S_n+\left(6 +34\frac{\delta m}{m}\right)\Sigma _n\right]
\Bigg)+\mathcal{O}\left(\frac{1}{c^8}\right) \;.
\end{align}

\end{subequations}

\section{Conclusions}
\label{sec:conclusion}

Building on previous work (Ref.~\cite{Marsat2012}, referred to as
Paper~I) we have presented complete results valid at the
next-to-next-to-leading order, corresponding to 3.5PN order, for the
spin-orbit contributions to the dynamics of spinning compact
binaries. We notably obtained the spin-orbit contribution to the
evolution (or precession) equations for the spins, the conserved
integrals of the motion associated with the Poincar\'e invariance of
the equations (energy, angular momentum and center-of-mass integral),
and the near-zone metric regularized at the location of the
particles. We also obtained the bulk metric
  (\textit{i.e.} at any field point in the near-zone) but with less
  precision in the $00$ and $0i$ components. The
results are given for general orbits in the frame of the binary's
center-of-mass, and reduced for quasi-circular orbits, which is the
mostly relevant case for realistic compact binaries detected by
  gravitational waves. We systematically make use of a precise
definition of the spin variables, which have a conserved Euclidean
norm.

The metric in the near zone is especially of interest in applications
such as the numerical study of a possible astrophysical environment
surrounding black-hole binaries with spins \cite{GNYC12}, and the
comparison with numerical computations of the gravitational self force
acting on a particle orbiting a Kerr black hole, which are based on
first (or second) order black hole perturbation theory
\cite{BBL12,FLS12}. More generally, all the present analytical
results (given explicitly as they are) should be useful for the
accurate comparisons between PN predictions and the full numerical
computations of the coalescence of two spinning black holes.

The next step in this current program will be to obtain the
next-to-next-to-leading order spin-orbit contributions in the
gravitational-wave phasing of spinning compact binaries. The GW
phasing including high-order spin effects (and especially high-order
spin-orbit effects) is of crucial importance for the parameter
estimation of signals received from compact binaries by the networks
of ground-based as well as, in the future, space-based GW detectors.

\section*{Acknowledgements}

A.B. is grateful for the support of the European Union FEDER
funds, the Spanish Ministry of Economy and Competitiveness project
FPA2010-16495 and the Conselleria d'Economia Hisenda i Innovacio of
the Govern de les Illes Balears.


%
%
\appendix

%
%

\section{Comparison with the ADM formalism}
\label{app:CompADM}

In this Appendix, we give the 3PN correspondence between our spin
variables and those of Ref.~\cite{Hartung2011}. The 2PN correspondence
had been investigated in \cite{Damour2008a}. In Section VII~D of
Paper~I, we showed that our results for the equations of motion were
equivalent to the ones obtained there in ADM variables, and we
constructed the extension of the contact transformation required to
relate the two methods, as well as the 2PN link between the ADM spin
variables and the harmonic-coordinates spin tensor. We shall use here
the same notations, with an overbar for all ADM quantities as well as
the convenient short-cut $\ov{\pi}\equiv \ov{p}/m$. Let us now extend
this correspondence in the dynamics to the 3PN spin precession
equations.

The ADM conserved-norm vector $\ovbf{S}$ obeys the precession
equation
\begin{equation}
	\frac{\ud \ovbf{S}}{\ud t} = \ovbf{\Omega} \times \ovbf{S} \, ,
\end{equation}
where the precession vector $\ovbf{\Omega}$ can be read directly on
the spin-orbit Hamiltonian $H_\mathrm{SO} = \ovbf{\Omega}_{1} \cdot
\ovbf{S}_{1} + \ovbf{\Omega}_{2} \cdot \ovbf{S}_{2}$ in
Ref.~\cite{Damour2008a}. The two conserved-norm spins start differing
at 2PN order, and from the fact that they must have the same magnitude
we may write
\begin{equation}
  \mathbf{S}(\ovbf{x},\ovbf{p}) = \ovbf{S} + 
  \bm{\theta}(\ovbf{x},\ovbf{p}) \times \ovbf{S} 
  + \mathcal{O}\left(\frac{1}{c^8}\right)\, ,
\end{equation}
where $\bm{\theta}$ represents a small rotation vector starting at
$\calO(1/c^4)$. Its expression is found by comparing the precession
equations for the two spin variables. Indeed, if we plug this relation into
the precession equation for $\mathbf{S}$, noticing that
$\mathbf{\Omega}(\ovbf{x},\ovbf{p}) = \ovbf{\Omega} + \calO(1/c^4)$, we
obtain
\begin{equation}
  \mathbf{\Omega} \times \ovbf{S} = \left( \ovbf{\Omega} + 
    \frac{\ud \bm{\theta}}{\ud t} + 
    \bm{\theta} \times \ovbf{\Omega} \right) \times \ovbf{S} + 
  \mathcal{O}\left(\frac{1}{c^8}\right)\, .
\end{equation}
Since we are looking for an expression of $\bm{\theta}$ that does not
feature any spin variable, we can forget about the degeneracy
remaining along the direction of $\ovbf{S}$ and we write
\begin{equation}
  \frac{\ud \bm{\theta}}{\ud t} + \bm{\theta} \times \ovbf{\Omega} = 
  \mathbf{\Omega}(\ovbf{x},\ovbf{p}) - \ovbf{\Omega} 
  + \mathcal{O}\left(\frac{1}{c^8}\right)\, .
\end{equation}
Using the method of undetermined coefficients for the dimensionless
vector $\bm{\theta}$, we obtain a unique solution (for particle 1),
\begin{subequations}
\begin{equation}
  \bm{\theta}_{1} = \frac{1}{c^4}
  \bm{\theta}_{1}^{\mathrm{2PN}} + 
  \frac{1}{c^6} \bm{\theta}_{1}^{\mathrm{3PN}} + \mathcal{O}\left(\frac{1}{c^8}\right)\, ,
\end{equation}
with the following expressions for the 2PN and 3PN
contributions:\footnote{Notice that the 2PN terms are different from equation
  (6.4) in \cite{Damour2008a}, since they compare the ADM spin variable to the
  different conserved variable $\mathbf{S}^{c}_\mathrm{BBF}$ used in
  Ref.~\cite{Blanchet2006}.}
\begin{align}
  \bm{\theta}_{1}^{\mathrm{2PN}} &= \frac{G m_{2}}{\ov{r}_{12}}
\left[-\frac{1}{4} \ovbf{n}_{12}\times\ovbm{\pi}_{1} (\ov{n}_{12}\ov{\pi}_{2})+ 
\ovbf{n}_{12}\times\ovbm{\pi}_{2} (\ov{n}_{12}\ov{\pi}_{2}) -
\frac{1}{4} \ovbm{\pi}_{1}\times\ovbm{\pi}_{2}  \right]\, ,\\
\bm{\theta}_{1}^{\mathrm{3PN}} &= 
\frac{G}{\ov{r}_{12}} {}^{(3)}\bm{\theta}_{1}^{0,1} m_{2} + 
\frac{G^{2}}{\ov{r}_{12}^{2}} 
\left[ {}^{(3)}\bm{\theta}_{1}^{1,1} m_{1}m_{2} + 
{}^{(3)}\bm{\theta}_{1}^{0,2} m_{2}^{2}\right]\,,\\
{}^{(3)}\bm{\theta}_{1}^{0,1} &= 
\ovbf{n}_{12}\times\ovbm{\pi}_{1} 
\left[ \frac{1}{16}\ov{\pi}_{1}^{2}(\ov{n}_{12}\ov{\pi}_{2}) + 
\frac{9}{16}(\ov{n}_{12}\ov{\pi}_{2})^{3} - 
\frac{5}{16}\ov{\pi}_{2}^{2}(\ov{n}_{12}\ov{\pi}_{2}) \right] \nn \\ 
	&+ \ovbf{n}_{12}\times\ovbm{\pi}_{2} 
\left[ -\frac{3}{4}(\ov{n}_{12}\ov{\pi}_{1})(\ov{n}_{12}\ov{\pi}_{2})^{2} - 
\frac{3}{4}(\ov{n}_{12}\ov{\pi}_{2})^{3} - 
\frac{1}{2}(\ov{n}_{12}\ov{\pi}_{2})(\ov{\pi}_{1}\ov{\pi}_{2}) \right. \nn \\ 
	& \qquad \qquad \quad \; \, \left. + 
\frac{1}{4}(\ov{n}_{12}\ov{\pi}_{1})\ov{\pi}_{2}^{2} + 
\frac{1}{4}(\ov{n}_{12}\ov{\pi}_{2})\ov{\pi}_{2}^{2} \right] \nn \\ 
	&+ \ovbm{\pi}_{1}\times\ovbm{\pi}_{2} 
\left[ \frac{1}{16}\ov{\pi}_{1}^{2} + 
\frac{13}{16}(\ov{n}_{12}\ov{\pi}_{2})^{2} + 
\frac{1}{2}(\ov{\pi}_{1}\ov{\pi}_{2}) - 
\frac{9}{16}\ov{\pi}_{2}^{2} \right] \;,\nn \\
	{}^{(3)}\bm{\theta}_{1}^{1,1} &= 
\ovbf{n}_{12}\times\ovbm{\pi}_{1} \left[ \frac{3}{8}(\ov{n}_{12}\ov{\pi}_{1})- 
(\ov{n}_{12}\ov{\pi}_{2}) \right] + \ovbf{n}_{12}\times\ovbm{\pi}_{2} 
\left[  -\frac{5}{2}(\ov{n}_{12}\ov{\pi}_{1}) +
\frac{1}{2}(\ov{n}_{12}\ov{\pi}_{2}) \right] + 
\frac{35}{16} \ovbm{\pi}_{1}\times\ovbm{\pi}_{2} \;,\nn \\
	{}^{(3)}\bm{\theta}_{1}^{0,2} &= 
\ovbf{n}_{12}\times\ovbm{\pi}_{1} 
\left[ -\frac{1}{8}(\ov{n}_{12}\ov{\pi}_{1}) + 
\frac{3}{2}(\ov{n}_{12}\ov{\pi}_{2}) \right] - 
\frac{41}{8}\ovbf{n}_{12}\times\ovbm{\pi}_{2} (\ov{n}_{12}\ov{\pi}_{2}) + 
\frac{3}{4} \ovbm{\pi}_{1}\times\ovbm{\pi}_{2}\;.
\end{align}
\end{subequations}
The existence of such a solution for $\bm{\theta}$ shows also that the
3PN precession equations are equivalent in the two formalisms, which
gives us further confidence in all results.

\section{Conversion from the conserved spin vector to the spin tensor}
\label{app:Explicitspinrelation}

In this Section we use the same notations for the orbital variables as
in Paper~I, except for the spin tensor, denoted here
  $\tilde{S}^{ij}$. From Eqs.~\eqref{eq:DefSmu} and~\eqref{eq:DefSi},
we obtain for our conserved norm spin vector, expressed in terms of
the spin tensor $\tilde{S}^{ij}$, the following expressions:
\begin{subequations} \label{eq:spin_conversion}
\begin{align}
	\mathbf{S}_{1} = \frac{1}{2} \varepsilon^{ijk}\tilde{S}_{1}^{jk} + 
\frac{1}{c^{2}}\mathbf{S}_{1}^{\mathrm{1PN}}   + 
\frac{1}{c^{4}}\mathbf{S}_{1}^{\mathrm{2PN}} + 
\frac{1}{c^{6}}\mathbf{S}_{1}^{\mathrm{3PN}} + \calO(1/c^8) \, ,
\end{align}
\begin{align}
	\mathbf{S}_{1}^{\mathrm{1PN}} =& {}^{(1)}\mathbf{S}_{1}^{0,0} + 
\frac{G}{r_{12}} {}^{(1)}\mathbf{S_{1}}^{0,1} m_{2} \, , \nn \\
	\mathbf{S}_{1}^{\mathrm{2PN}} =& {}^{(2)}\mathbf{S}_{1}^{0,0} + 
\frac{G}{r_{12}} {}^{(2)}\mathbf{S}_{1}^{0,1} m_{2} + 
\frac{G^{2}}{r_{12}^{2}} \left[  {}^{(2)}\mathbf{S}_{1}^{1,1}
  m_{1}m_{2} + 
{}^{(2)}\mathbf{S}_{1}^{0,2} m_{2}^{2} \right] \, ,  \nn \\
	\mathbf{S_{1}}^{\mathrm{3PN}} =& {}^{(3)}\mathbf{S_{1}}^{0,0} + 
\frac{G}{r_{12}} {}^{(3)}\mathbf{S}_{1}^{0,1} m_{2} + 
\frac{G^{2}}{r_{12}^{2}} \left[  {}^{(3)}\mathbf{S}_{1}^{1,1}
  m_{1}m_{2} + 
{}^{(3)}\mathbf{S}_{1}^{0,2} m_{2}^{2} \right]  \nn \\
	& +  \frac{G^{3}}{r_{12}^{3}} \left[
      {}^{(3)}\mathbf{S}_{1}^{2,1} m_{1}^{2}m_{2} +
      {}^{(3)}\mathbf{S}_{1}^{1,2} m_{1}m_{2}^{2} + 
{}^{(3)}\mathbf{S}_{1}^{0,3} m_{2}^{3} \right] \, ,
\end{align}
\end{subequations}
with, using like in Paper~I the notation $(\varepsilon a \tilde{S}) =
\varepsilon^{ijk}a^{i}\tilde{S}^{jk}$: 
%
\begin{align}
  \left({}^{(1)}S_{1}^{0,0}\right)^{i} &=
  \frac{1}{4} v_{1}^{2} \varepsilon^{ijk}\tilde{S}_{1}^{jk} - \frac{1}{4} (\varepsilon{}v_{1}\tilde{S}_{1})v_{1}^{i} - \varepsilon^{ijk}v_{1}^{j}v_{1}^{l}\tilde{S}_{1}^{lk} \;,\nonumber \\
  \left({}^{(1)}S_{1}^{0,1}\right)^{i} &=
  \varepsilon^{ijk}\tilde{S}_{1}^{jk} \;,\nonumber \\
  \left({}^{(2)}S_{1}^{0,0}\right)^{i} &= \frac{3}{16} v_{1}^{4}
  \varepsilon^{ijk}\tilde{S}_{1}^{jk} - \frac{3}{16} v_{1}^{2}
  (\varepsilon{}v_{1}\tilde{S}_{1})v_{1}^{i} - \frac{1}{2} v_{1}^{2}
  \varepsilon^{ijk}v_{1}^{j}v_{1}^{l}
  \tilde{S}_{1}^{lk} \;,\nonumber \\
  \left({}^{(2)}S_{1}^{0,1}\right)^{i} &= \varepsilon^{ijk}
  \tilde{S}_{1}^{jk} \left[ -\frac{1}{2} (n_{12}v_{2})^2 +
    \frac{3}{2} v_{1}^{2} - 2 (v_{1}v_{2}) + v_{2}^{2} \right] - \frac{3}{2} (\varepsilon{}v_{1}\tilde{S}_{1})v_{1}^{i} + (\varepsilon{}v_{1}\tilde{S}_{1})v_{2}^{i} \nonumber \\
  & + (\varepsilon{}v_{2}\tilde{S}_{1})v_{1}^{i} - (\varepsilon{}v_{2}\tilde{S}_{1})v_{2}^{i} - 6 \varepsilon^{ijk}v_{1}^{j}v_{1}^{l}\tilde{S}_{1}^{lk} + 4 \varepsilon^{ijk}v_{1}^{j}v_{2}^{l}\tilde{S}_{1}^{lk} \;,\nonumber \\
  \left({}^{(2)}S_{1}^{1,1}\right)^{i} &= -
  \frac{3}{2} \varepsilon^{ijk}\tilde{S}_{1}^{jk} + 2 (\varepsilon{}n_{12}\tilde{S}_{1})n_{12}^{i} \;,\nonumber \\
  \left({}^{(2)}S_{1}^{0,2}\right)^{i} &=
  \frac{3}{4} \varepsilon^{ijk}\tilde{S}_{1}^{jk} - \frac{1}{4} (\varepsilon{}n_{12}\tilde{S}_{1})n_{12}^{i} \;,\nonumber \\
  \left({}^{(3)}S_{1}^{0,0}\right)^{i} &=
  \frac{5}{32} v_{1}^{6} \varepsilon^{ijk}\tilde{S}_{1}^{jk} - \frac{5}{32} v_{1}^{4} (\varepsilon{}v_{1}\tilde{S}_{1})v_{1}^{i} - \frac{3}{8} v_{1}^{4} \varepsilon^{ijk}v_{1}^{j}v_{1}^{l}\tilde{S}_{1}^{lk} \;,\nonumber \\
  \left({}^{(3)}S_{1}^{0,1}\right)^{i} &=
  \varepsilon^{ijk}\tilde{S}_{1}^{jk} \left[ \frac{3}{8}
    (n_{12}v_{2})^4 - \frac{3}{4} (n_{12}v_{2})^2 v_{1}^{2} +
    \frac{15}{8} v_{1}^{4} + (n_{12}v_{2})^2 (v_{1}v_{2}) - 3
    v_{1}^{2} (v_{1}v_{2}) +
    (v_{1}v_{2})^2 \right. \nonumber \\
  & \quad \left. - (n_{12}v_{2})^2 v_{2}^{2} + \frac{3}{2} v_{1}^{2}
    v_{2}^{2} - 2 (v_{1}v_{2}) v_{2}^{2} + v_{2}^{4} \right] +
  (\varepsilon{}v_{2}\tilde{S}_{1})v_{2}^{i} \left[ \frac{1}{2}
    (n_{12}v_{2})^2 - \frac{1}{2} v_{1}^{2} -
    v_{2}^{2} \right] \nonumber \\
  & + (\varepsilon{}v_{1}\tilde{S}_{1})v_{1}^{i} \left[ \frac{3}{4}
    (n_{12}v_{2})^2 - \frac{15}{8} v_{1}^{2} + \frac{3}{2}
    (v_{1}v_{2}) - \frac{3}{2} v_{2}^{2} \right] +
  (\varepsilon{}v_{1}\tilde{S}_{1})v_{2}^{i}
  \left[ -\frac{1}{2} (n_{12}v_{2})^2 + \frac{3}{4} v_{1}^{2} \right. \nonumber \\
  & \quad \left. - \frac{1}{4} (v_{1}v_{2}) + v_{2}^{2} \right] +
  (\varepsilon{}v_{2}\tilde{S}_{1})v_{1}^{i} \left[ -\frac{1}{2}
    (n_{12}v_{2})^2 + \frac{3}{4} v_{1}^{2} -
    \frac{1}{4} (v_{1}v_{2}) + v_{2}^{2} \right] \nonumber \\
  & + \varepsilon^{ijk}v_{1}^{j}v_{1}^{l}\tilde{S}_{1}^{lk} \left[ 3
    (n_{12}v_{2})^2 - 5 v_{1}^{2} + 8 (v_{1}v_{2}) - 6 v_{2}^{2}
    \vphantom{\frac{1}{2}} \right] + 2
  (\varepsilon^{jkl}v_{1}^{j}v_{2}^{k}v_{1}^{m}\tilde{S}_{1}^{ml})
  v_{1}^{i} \nonumber \\
  & + \varepsilon^{ijk}v_{1}^{j}v_{2}^{l}\tilde{S}_{1}^{lk} \left[ -2
    (n_{12}v_{2})^2 + 2 v_{1}^{2} - 4 (v_{1}v_{2}) + 4 v_{2}^{2}
    \vphantom{\frac{1}{2}} \right] - 2
  (\varepsilon^{jkl}v_{1}^{j}v_{2}^{k}v_{1}^{m}\tilde{S}_{1}^{ml})
  v_{2}^{i} \;,\nonumber \\
  \left({}^{(3)}S_{1}^{1,1}\right)^{i} &=
  \varepsilon^{ijk}\tilde{S}_{1}^{jk} \left[ \frac{1}{8}
    (n_{12}v_{1})^2 + \frac{7}{4} (n_{12}v_{1}) (n_{12}v_{2}) +
    \frac{1}{8} (n_{12}v_{2})^2 - \frac{3}{8} v_{1}^{2} - \frac{3}{4}
    (v_{1}v_{2}) +
    \frac{3}{8} v_{2}^{2} \right] \nonumber \\
  & + (\varepsilon{}n_{12}\tilde{S}_{1})n_{12}^{i} \left[ 4
    (n_{12}v_{1})^2 - 8 (n_{12}v_{1}) (n_{12}v_{2}) +
    4 v_{1}^{2} - 6 (v_{1}v_{2}) + 3 v_{2}^{2} \vphantom{\frac{1}{2}} \right] \nonumber \\
  & + (\varepsilon{}n_{12}\tilde{S}_{1})v_{1}^{i} \left[ -7
    (n_{12}v_{1}) + \frac{9}{2} (n_{12}v_{2}) \right] +
  (\varepsilon{}n_{12}\tilde{S}_{1})v_{2}^{i} \left[ \frac{15}{2}
    (n_{12}v_{1}) -
    \frac{9}{2} (n_{12}v_{2}) \right] \nonumber \\
  & + (\varepsilon{}v_{1}\tilde{S}_{1})n_{12}^{i} \left[ -7
    (n_{12}v_{1}) + \frac{9}{2} (n_{12}v_{2}) \right] +
  (\varepsilon{}v_{2}\tilde{S}_{1})n_{12}^{i} \left[ \frac{15}{2}
    (n_{12}v_{1}) -
    \frac{9}{2} (n_{12}v_{2}) \right] \nonumber \\
  & + \frac{13}{4} (\varepsilon{}v_{1}\tilde{S}_{1})v_{1}^{i} -
  \frac{5}{2} (\varepsilon{}v_{1}\tilde{S}_{1})v_{2}^{i} - \frac{5}{2}
  (\varepsilon{}v_{2}\tilde{S}_{1})v_{1}^{i} + \frac{5}{2}
  (\varepsilon{}v_{2}\tilde{S}_{1})v_{2}^{i} - 4
  (\varepsilon^{jkl}n_{12}^{j}v_{1}^{k}v_{1}^{m}\tilde{S}_{1}^{ml})
  n_{12}^{i} \nonumber \\
  & + \varepsilon^{ijk}v_{1}^{j}n_{12}^{l}\tilde{S}_{1}^{lk}
  \left[ -2 (n_{12}v_{1}) - 2 (n_{12}v_{2}) \vphantom{\frac{1}{2}} \right] + \varepsilon^{ijk}v_{1}^{j}v_{1}^{l}\tilde{S}_{1}^{lk} + 2 \varepsilon^{ijk}v_{1}^{j}v_{2}^{l}\tilde{S}_{1}^{lk} \;,\nonumber \\
  \left({}^{(3)}S_{1}^{0,2}\right)^{i} &=
  \varepsilon^{ijk}\tilde{S}_{1}^{jk} \left[ \frac{1}{4}
    (n_{12}v_{1})^2 - \frac{1}{2} (n_{12}v_{1}) (n_{12}v_{2}) -
    \frac{1}{2} (n_{12}v_{2})^2 + \frac{33}{8} v_{1}^{2} -
    \frac{15}{2} (v_{1}v_{2}) + \frac{15}{4} v_{2}^{2} \right] \nonumber \\
  & + (\varepsilon{}n_{12}\tilde{S}_{1})n_{12}^{i} \left[ \frac{1}{2}
    (n_{12}v_{2})^2 - \frac{1}{8} v_{1}^{2} \right] +
  (\varepsilon{}n_{12}\tilde{S}_{1})v_{1}^{i} \left[ -\frac{1}{16}
    (n_{12}v_{1}) +
    \frac{1}{4} (n_{12}v_{2}) \right] \nonumber \\
  & - \frac{1}{4} (n_{12}v_{2})
  (\varepsilon{}n_{12}\tilde{S}_{1})v_{2}^{i} +
  (\varepsilon{}v_{1}\tilde{S}_{1})n_{12}^{i} \left[ -\frac{1}{16}
    (n_{12}v_{1}) +
    \frac{1}{4} (n_{12}v_{2}) \right] - \frac{33}{8} (\varepsilon{}v_{1}\tilde{S}_{1})v_{1}^{i} \nonumber \\
  & + \frac{15}{4} (\varepsilon{}v_{1}\tilde{S}_{1})v_{2}^{i} -
  \frac{1}{4} (n_{12}v_{2})
  (\varepsilon{}v_{2}\tilde{S}_{1})n_{12}^{i} + \frac{15}{4} (\varepsilon{}v_{2}\tilde{S}_{1})v_{1}^{i} - \frac{15}{4} (\varepsilon{}v_{2}\tilde{S}_{1})v_{2}^{i} + 15 \varepsilon^{ijk}v_{1}^{j}v_{2}^{l}\tilde{S}_{1}^{lk} \nonumber \\
  & + \frac{1}{2} (\varepsilon^{jkl}n_{12}^{j}v_{1}^{k}v_{1}^{m}
  \tilde{S}_{1}^{ml}) n_{12}^{i} +
  \varepsilon^{ijk}v_{1}^{j}n_{12}^{l}\tilde{S}_{1}^{lk} \left[ -
    (n_{12}v_{1}) + (n_{12}v_{2}) \vphantom{\frac{1}{2}} \right] -
  \frac{33}{2} \varepsilon^{ijk}v_{1}^{j}v_{1}^{l}
  \tilde{S}_{1}^{lk} \;, \nonumber \\
  \left({}^{(3)}S_{1}^{2,1}\right)^{i} &=
  \frac{5}{2} \varepsilon^{ijk}\tilde{S}_{1}^{jk} - 7 (\varepsilon{}n_{12}\tilde{S}_{1})n_{12}^{i} \;,\nonumber \\
  \left({}^{(3)}S_{1}^{1,2}\right)^{i} &=
  \frac{17}{6} \varepsilon^{ijk}\tilde{S}_{1}^{jk} - \frac{3}{4} (\varepsilon{}n_{12}\tilde{S}_{1})n_{12}^{i} \;, \nonumber \\
  \left({}^{(3)}S_{1}^{0,3}\right)^{i} &= \frac{1}{2}
  \varepsilon^{ijk}\tilde{S}_{1}^{jk} - \frac{1}{2}
  (\varepsilon{}n_{12}\tilde{S}_{1})n_{12}^{i} \, .
\end{align}
As stated above, these conserved spin variables turn out to be the
same as the conserved spin vectors of \cite{Blanchet2006}, at 2PN
order, when evaluated in the center-of-mass frame.

\section{Center-of-mass position}
\label{appendix:CMreduction}

In a general frame, the expressions of the whole set of conserved
integrals of motion are too lengthy and we therefore chose to present
them directly reduced to the center-of-mass frame. However, in this
Appendix, we provide the complete expression for the center-of-mass
position $\mathbf{G}$ (necessarily in a general frame) since
this might be used in the future to reduce higher order results to the
center of mass. Denoting the PN expansion of $\mathbf{G}$ in the form
\begin{subequations}
\begin{align}\label{Gistruct}
\mathbf{G} &=
\mathbf{G}_\mathrm{N}+\frac{1}{c^2}\mathbf{G}_\mathrm{1PN}+\frac{1}{c^3}
\mathop{\mathbf{G}}_{S}{}_{\!\mathrm{1.5PN}}
+\frac{1}{c^4}\left[\mathbf{G}_\mathrm{2PN}+
\mathop{\mathbf{G}}_{SS}{}_{\!\mathrm{2PN}}\right]
+\frac{1}{c^5} \mathop{\mathbf{G}}_{S}{}_{\!\mathrm{2.5PN}} \nonumber\\
& \qquad + \frac{1}{c^6}\left[\mathbf{G}_\mathrm{3PN}+
\mathop{\mathbf{G}}_{SS}{}_{\!\mathrm{3PN}}\right] + 
\frac{1}{c^7} \mathop{\mathbf{G}}_{S}{}_{\!\mathrm{3.5PN}} + 
\mathcal{O}\left(\frac{1}{c^8}\right) \, ,
\end{align}
we have for the leading and next-to-leading order spin-orbit
contributions
\begin{align}
	\mathop{\mathbf{G}}_{S}{}_{\!\mathrm{1.5PN}} &= -
\mathbf{ S_ {1} \times v_{1}}+ 1\leftrightarrow2 \, ,\\
	\mathop{\mathbf{G}}_{S}{}_{\!\mathrm{2.5PN}} &= {}^{(2.5)}\mathbf{g}_{0,0}+
\frac{G}{r_{12}} {}^{(2.5)}\mathbf{g}_{0,1} m_{2}+ 1\leftrightarrow2 \, ,
\end{align}
with 
\allowdisplaybreaks{
\begin{align}
    {}^{(2.5)}\mathbf{g}_{0,0} &= -\frac{1}{2} v_{1}^{2} \, \mathbf{S}_{1} \times \mathbf{v}_{1} \nonumber \\
    {}^{(2.5)}\mathbf{g}_{0,1} &= \frac{3}{4} (n_{12},S_{1},v_{1}) \, \mathbf{n}_{12} - (n_{12},S_{1},v_{2}) \, \mathbf{n}_{12} - \frac{3}{4} (n_{12},S_{1},v_{1}) \; \mathbf{\tilde{y}}_1 + 2 (n_{12},S_{1},v_{2}) \; \mathbf{\tilde{y}}_1 \;, \nonumber \\
& - \frac{1}{4} (n_{12},S_{1},v_{1}) \, \mathbf{\tilde{y}}_2 - 2 (n_{12},S_{1},v_{2}) \, \mathbf{\tilde{y}}_2 - 2 \, \mathbf{S}_{1} \times \mathbf{v}_{1} + 2 \, \mathbf{S}_{1} \times \mathbf{v}_{2} \nonumber \\
& + \, \mathbf{ n}_{12} \times \mathbf{S}_{1} \left[ - (n_{12}v_{1}) -  (n_{12}v_{2}) \right]\;,
\end{align}}\noindent 
where we have defined for convenience
$\tilde{\mathbf{y}}_A=\mathbf{y}_A/r_{12}$. Let us emphasize again
that these two contributions are the only ones required to derive
Eq.~\eqref{y1CM}, \textit{i.e.} to perform our center of mass
reduction. In addition to these, the next-to-next-to-leading
correction is given by
\begin{align}
	\mathop{\mathbf{G}}_{S}{}_{\!\mathrm{3.5PN}} &= {}^{(3.5)}\mathbf{g}_{0,0} +
\frac{G}{r_{12}}{}^{(3.5)}\mathbf{g}_{0,1} m_{2} + \frac{G^{2}}{r_{12}^{2}} 
\left[ {}^{(3.5)}\mathbf{g}_{0,2} m_{2}^{2} + 
{}^{(3.5)}\mathbf{g}_{1,1} m_{1}m_{2} \right] + 1\leftrightarrow2 \, ,
\end{align}
\allowdisplaybreaks{
\begin{align}
    {}^{(3.5)}\mathbf{g}_{0,0} &= -\frac{3}{8} v_{1}^{4} \mathbf{S}_{1} \times \mathbf{v}_{1} \;, \nonumber \\
    {}^{(3.5)}\mathbf{g}_{0,1} &= (n_{12},S_{1},v_{1}) \mathbf{n}_{12} \left[ -\frac{3}{4} (n_{12}v_{1})^2 - \frac{9}{8} (n_{12}v_{1}) (n_{12}v_{2}) - \frac{3}{4} (n_{12}v_{2})^2 + \frac{9}{16} v_{1}^{2} - \frac{7}{8} v_{2}^{2} \right] \nonumber \\
& + (n_{12},S_{1},v_{2}) \mathbf{n}_{12} \! \! \left[ \frac{3}{4} (n_{12}v_{1})^2 + \frac{3}{2} (n_{12}v_{1}) (n_{12}v_{2}) - \frac{3}{4} (n_{12}v_{2})^2 - \frac{3}{8} v_{1}^{2} + \frac{3}{2} (v_{1}v_{2}) - \frac{1}{4} v_{2}^{2} \right] \nonumber \\
& + (S_{1},v_{1},v_{2}) \, \mathbf{n}_{12} \left[ \frac{1}{4} (n_{12}v_{1}) + 2 (n_{12}v_{2}) \right] + (n_{12},S_{1},v_{1}) \; \mathbf{v}_1 \left[ - (n_{12}v_{1}) - \frac{9}{8} (n_{12}v_{2}) \right] \nonumber \\
& + (n_{12},S_{1},v_{2}) \; \mathbf{v}_1 \left[ \frac{5}{8} (n_{12}v_{1}) + \frac{1}{2} (n_{12}v_{2}) \right] + (n_{12},S_{1},v_{1}) \, \mathbf{v}_2 \left[ \frac{3}{2} (n_{12}v_{1}) + \frac{5}{2} (n_{12}v_{2}) \right] \nonumber \\
& + (n_{12},S_{1},v_{2}) \, \mathbf{v}_2 \left[ -\frac{3}{2} (n_{12}v_{1}) - \frac{1}{2} (n_{12}v_{2}) \right] + \frac{9}{8} (S_{1},v_{1},v_{2}) \; \mathbf{v}_1 - (S_{1},v_{1},v_{2}) \, \mathbf{v}_2 \nonumber \\
& + (n_{12},S_{1},v_{1}) \; \mathbf{\tilde{y}}_1 \left[ -\frac{3}{8} (n_{12}v_{1}) (n_{12}v_{2}) - \frac{5}{16} v_{1}^{2} + \frac{1}{8} (v_{1}v_{2}) - \frac{7}{8} v_{2}^{2} \right] + v_{2}^{2} (n_{12},S_{1},v_{2}) \; \mathbf{\tilde{y}}_1 \nonumber \\
& + (n_{12},S_{1},v_{1}) \, \mathbf{\tilde{y}}_2 \left[ -\frac{21}{8} (n_{12}v_{1}) (n_{12}v_{2}) - \frac{3}{2} (n_{12}v_{2})^2 - \frac{3}{16} v_{1}^{2} - \frac{9}{8} (v_{1}v_{2}) + \frac{15}{8} v_{2}^{2} \right] \nonumber \\
& + (n_{12},S_{1},v_{2}) \, \mathbf{\tilde{y}}_2 \left[ 3 (n_{12}v_{1})^2 + 3 (n_{12}v_{1}) (n_{12}v_{2}) -  v_{1}^{2} + 3 (v_{1}v_{2}) - 3 v_{2}^{2} \right] \nonumber \\
& + (S_{1},v_{1},v_{2}) \, \mathbf{\tilde{y}}_2 \left[ \frac{15}{8} (n_{12}v_{1}) + 2 (n_{12}v_{2}) \right] - \frac{7}{8} (n_{12}v_{1}) (S_{1},v_{1},v_{2}) \; \mathbf{\tilde{y}}_1 \nonumber \\
& + \, \mathbf{S}_{1} \times \mathbf{v}_{1} \left[ \frac{1}{2} (n_{12}v_{1})^2 + \frac{5}{8} (n_{12}v_{1}) (n_{12}v_{2}) - 3 v_{1}^{2} + \frac{23}{8} (v_{1}v_{2}) -  v_{2}^{2} \right] \nonumber \\
& + \, \mathbf{S}_{1} \times \mathbf{v}_{2} \left[ -\frac{1}{8} (n_{12}v_{1})^2 + (n_{12}v_{1}) (n_{12}v_{2}) -  (n_{12}v_{2})^2 + \frac{17}{8} v_{1}^{2} - 3 (v_{1}v_{2}) + 2 v_{2}^{2} \right] \nonumber \\
& + \, \mathbf{ n}_{12} \times \mathbf{S}_{1} \left[ \frac{3}{4} (n_{12}v_{1})^3 + \frac{3}{4} (n_{12}v_{1})^2 (n_{12}v_{2}) + \frac{3}{4} (n_{12}v_{1}) (n_{12}v_{2})^2 + \frac{3}{4} (n_{12}v_{2})^3 \right. \nonumber \\
& \left. \quad - \frac{3}{4} (n_{12}v_{1}) v_{1}^{2} - \frac{1}{8} (n_{12}v_{2}) v_{1}^{2} + \frac{3}{8} (n_{12}v_{1}) (v_{1}v_{2}) - \frac{1}{2} (n_{12}v_{2}) (v_{1}v_{2}) + \frac{1}{4} (n_{12}v_{1}) v_{2}^{2}  \right. \nonumber \\
& \left. \quad - \frac{5}{4} (n_{12}v_{2}) v_{2}^{2} \right] \;, \nonumber \\
    {}^{(3.5)}\mathbf{g}_{1,1} &= -\frac{39}{4} (n_{12},S_{1},v_{1}) \, \mathbf{n}_{12} + 13 (n_{12},S_{1},v_{2}) \, \mathbf{n}_{12} - \frac{9}{4} (n_{12},S_{1},v_{1}) \; \mathbf{\tilde{y}}_1 +  (n_{12},S_{1},v_{2}) \; \mathbf{\tilde{y}}_1 \nonumber \\
& + \frac{17}{4} (n_{12},S_{1},v_{1}) \, \mathbf{\tilde{y}}_2 - (n_{12},S_{1},v_{2}) \, \mathbf{\tilde{y}}_2 + 10 \, \mathbf{S}_{1} \times \mathbf{v}_{1} - 10 \, \mathbf{S}_{1} \times \mathbf{v}_{2} \nonumber \\
& + \, \mathbf{ n}_{12} \times \mathbf{S}_{1} \left[ \frac{29}{2} (n_{12}v_{1}) - \frac{23}{2} (n_{12}v_{2}) \right] \;,\nonumber \nonumber \\
    {}^{(3.5)}\mathbf{g}_{0,2} &= \frac{55}{8} (n_{12},S_{1},v_{1}) \, \mathbf{n}_{12} - \frac{19}{4} (n_{12},S_{1},v_{2}) \, \mathbf{n}_{12} + \frac{1}{2} (n_{12},S_{1},v_{1}) \; \mathbf{\tilde{y}}_1 - \frac{25}{8} (n_{12},S_{1},v_{2}) \; \mathbf{\tilde{y}}_1 \nonumber \\
& - \frac{1}{2} (n_{12},S_{1},v_{1}) \, \mathbf{\tilde{y}}_2 + \frac{49}{8} (n_{12},S_{1},v_{2}) \, \mathbf{\tilde{y}}_2 - \frac{45}{8} \, \mathbf{S}_{1} \times \mathbf{v}_{1} + \frac{45}{8} \, \mathbf{S}_{1} \times \mathbf{v}_{2} \nonumber \\
& + \, \mathbf{ n}_{12} \times \mathbf{S}_{1} \left[ -\frac{59}{8} (n_{12}v_{1}) + \frac{87}{8} (n_{12}v_{2}) \right] \;.
\end{align}}
\end{subequations}

This result translates into a next-to-next-to-leading order correction
to the vector $\mathbf{z}$ in Eqs.~\eqref{y1CM}, of the form
\begin{subequations}

\begin{align}
\mathbf{\delta z}=\frac{1}{c^7}\left[{}^{(3.5)}\mathbf{z}^{(0)}+ 
\frac{G m}{r}{}^{(3.5)}\mathbf{z}^{(1)}+
\frac{G^2 m^2}{r^2}{}^{(3.5)}\mathbf{z}^{(2)}\right]\;,
\end{align}
with
\allowdisplaybreaks{
\begin{align}
   {}^{(3.5)} \mathbf{z}^{(0)} &= \, \mathbf{ \Sigma \times v } \left(-\frac{3}{8} + \frac{13}{4} \nu -7 \nu^2\right) v^{4}\;, \nonumber \\ 
   {}^{(3.5)} \mathbf{z}^{(1)} &= (n,\Sigma ,v) \, \mathbf{n} \left[\left(-\frac{3}{4} + \frac{9}{2} \nu -6 \nu^2\right) (nv)^2+\left(\frac{1}{4} + \nu -8 \nu^2\right) v^{2}\right] \nonumber \\
& + (n,S,v) \, \mathbf{n} \left[\frac{\delta m}{m}\left(-\frac{3}{2} + \frac{9}{2} \nu\right) (nv)^2+\frac{\delta m}{m}\left(3 - \nu\right) v^{2}\right] \nonumber \\
& + (n,\Sigma ,v) \, \mathbf{v} \left(-1 + \frac{7}{2} \nu + 2 \nu^2\right) (nv) + (n,S,v) \, \mathbf{v} \frac{\delta m}{m}\left(-\frac{3}{2} - \nu\right) (nv) \nonumber \\
& + \, \mathbf{ n \times S } \left[\frac{\delta m}{m}\left(\frac{9}{8} -\frac{9}{4} \nu\right) (nv)^3+\frac{\delta m}{m}\left(-\frac{7}{8} + 4 \nu\right) (nv) v^{2}\right] \nonumber \\
& + \, \mathbf{ n \times \Sigma } \left[\left(\frac{3}{4} -\frac{9}{2} \nu + 6 \nu^2\right) (nv)^3+\left(-\frac{3}{4} + \frac{23}{4} \nu -11 \nu^2\right) (nv) v^{2}\right] \nonumber \\
& + \, \mathbf{ S \times v } \left[\frac{\delta m}{m}\left(\frac{9}{8} -\frac{3}{4} \nu\right) (nv)^2+\frac{\delta m}{m}\left(-\frac{3}{8} + \nu\right) v^{2}\right] \nonumber \\
& + \, \mathbf{ \Sigma \times v } \left[\left(\frac{1}{2} -2 \nu + \frac{9}{2} \nu^2\right) (nv)^2+\left(-3 + \frac{25}{4} \nu + 6 \nu^2\right) v^{2}\right] \;, \nonumber \\ 
    {}^{(3.5)}\mathbf{z}^{(2)} &= (n,\Sigma ,v) \, \mathbf{n} \left(\frac{59}{8} -\frac{161}{4} \nu + 6 \nu^2\right) + \frac{171}{8}\frac{\delta m}{m} (n,S,v) \, \mathbf{n} \nonumber \\
& + \, \mathbf{ n \times S } \frac{\delta m}{m}\left(-\frac{133}{8} -4 \nu\right) (nv) + \, \mathbf{ n \times \Sigma } \left(-\frac{59}{8} + \frac{137}{4} \nu + 12 \nu^2\right) (nv) \nonumber \\
& + \, \mathbf{ S \times v } \frac{\delta m}{m}\left(-\frac{97}{8} - \nu\right) + \, \mathbf{ \Sigma \times v } \left(-\frac{45}{8} + \frac{119}{4} \nu - \nu^2\right)\;.
\end{align}}
\end{subequations}

To conclude this Appendix, let us show explicitly that the above
$\mathcal{O}(1/c^7)$ correction to $\mathbf{y}_1$ written in terms of
the CM variables cancels out when reducing our general frame results
to the center-of-mass frame. Let us denote by $\mathbf{z}$ the
$\mathcal{O}(1/c^2)$ correction to the Newtonian expressions of
$\mathbf{y}_1$ and $\mathbf{y}_2$, as defined by
Eqs.~\eqref{y1CM}. Since we want to obtain expressions up to
$\mathcal{O}(1/c^7)$ in the CM frame, the terms of order
$\mathcal{O}(1/c^7)$ in $\mathbf{z}$ can only play a role in the
reduction of the Newtonian parts. These Newtonian contributions are
very simple, so we can check that the expression of $\mathbf{z}$ is
needed only up to $\mathcal{O}(1/c^5)$.

The simplest case is that of the relative Newtonian acceleration
$\mathbf{a}_\mathrm{N}=-G m \,\mathbf{n_{12}}/r_{12}^2$, which
receives no higher order correction when reexpressing it in the CM
frame. The same argument applies to the near zone metric. By contrast,
the Newtonian energy
\begin{equation}
E_\mathrm{N} = \frac{1}{2}m \nu v^2+ 
m \left(\frac{\ud\mathbf{z}}{\ud t}\right)^2-
\frac{G m \nu}{2 r}\,,
\end{equation}
%
%
receives a PN correction when expressed in terms of the CM
quantities. However, since the terms linear in $\mathbf{z}$ cancel out
while $\ud\mathbf{z}/\ud t=\mathcal{O}(1/c^2)$, we only need
$\mathbf{z}$ up to the order $\mathcal{O}(1/c^5)$ to get the energy up
to the 3.5PN level. A similar argument applies to the Newtonian
angular momentum, which is deprived of linear-in-$\mathbf{z}$
contributions:
\begin{equation}
\mathbf{L}_\mathrm{N} = m \nu \,\mathbf{x} \times \mathbf{v} + 
m\, \mathbf{z} \times \frac{\ud\mathbf{z}}{\ud t}\,.
\end{equation}
%

\section{Orbital angular momentum}
\label{app:angmom}

The conserved total angular momentum (such that $\ud \mathbf{J}/\ud t
= 0$) is conventionaly split into an orbital angular momentum, which
will also receive contributions from the spins, and the sum of spins
of the system:
\begin{align}
\mathbf{J}=\mathbf{L} +\frac{\mathbf{S} }{c}\,. 
\end{align}
This actually represents the definition of the orbital angular
momentum $\mathbf{L}$, which depends on the choice of variables for
the individual spins. With this definition, the structure of the
orbital part is
\begin{align}\label{ListructCM}
\mathbf{L} &=\nu\bigg\{
\boldsymbol\ell_\mathrm{N}+\frac{1}{c^2}\boldsymbol\ell_\mathrm{1PN}+\frac{1}{c^3}
\mathop{\boldsymbol\ell}_{S} {}_{\!\mathrm{1.5PN}}
+\frac{1}{c^4}\left[\boldsymbol\ell_\mathrm{2PN}+
\mathop{\boldsymbol\ell}_{SS}{}_{\!\mathrm{2PN}}\right]
+\frac{1}{c^5} \mathop{\boldsymbol\ell}_{S}{}_{\!\mathrm{2.5PN}} \nonumber\\
& \qquad + \frac{1}{c^6}\left[\boldsymbol\ell_\mathrm{3PN}+
\mathop{\boldsymbol\ell}_{SS}{}_{\!\mathrm{3PN}}\right] + 
\frac{1}{c^7} \mathop{\boldsymbol\ell}_{S}{}_{\!\mathrm{3.5PN}} + 
\mathcal{O}\left(\frac{1}{c^8}\right)\bigg\} \,.
\end{align}
The non-spin terms up to 3PN order can be found in
Ref.~\cite{Blanchet2003}. Again neglecting the spin-spin terms, we
arrive at the following spin-orbit contributions. At leading and
next-to-leading order, we have
\begin{subequations}
\begin{align}
\mathop{\bm{\ell}}_{S}{}_{\!\mathrm{1.5PN}}={}^{(1.5)}\bm{\ell}_0+
{}^{(1.5)}\bm{\ell}_1 \frac{G m}{r}\, ,
\end{align}
%
\allowdisplaybreaks{
\begin{align}
    {}^{(1.5)}\mathbf{\boldsymbol\ell}_{0} &= -\frac{1}{2}\frac{\delta m}{m} (\Sigma v) \, \mathbf{v} - \frac{1}{2} (Sv) \, \mathbf{v} + \frac{1}{2} v^{2} \, \mathbf{S} + \frac{1}{2}\frac{\delta m}{m} v^{2} \, \mathbf{\Sigma} \;, \nonumber \\ 
    {}^{(1.5)}\mathbf{\boldsymbol\ell}_{1} &= \frac{\delta m}{m} (n\Sigma ) \, \mathbf{n} + 3 (nS) \, \mathbf{n} - 3 \, \mathbf{S} - \frac{\delta m}{m} \, \mathbf{\Sigma} \;,
\end{align}}\noindent
\end{subequations}
%
\begin{subequations}
\begin{align}
\mathop{\boldsymbol\ell}_{S}{}_{\!\mathrm{2.5PN}}=
{}^{(2.5)}\mathbf{\boldsymbol\ell}_0
+{}^{(2.5)}\mathbf{\boldsymbol\ell}_1 \frac{G m}{r}
+{}^{(2.5)}\mathbf{\boldsymbol\ell}_2 \frac{G^2 m^2}{r^2} \, ,
\end{align}
%
\allowdisplaybreaks{
\begin{align}
    {}^{(2.5)}\mathbf{\boldsymbol\ell}_{0} &= (\Sigma v) \, \mathbf{v} \frac{\delta m}{m}\left(-\frac{3}{8} + \frac{5}{4} \nu\right) v^{2} + (Sv) \, \mathbf{v} \left(-\frac{3}{8} + \frac{9}{8} \nu\right) v^{2} + \, \mathbf{S} \left(\frac{3}{8} -\frac{9}{8} \nu\right) v^{4}  \nonumber \\
& + \, \mathbf{\Sigma} \frac{\delta m}{m}\left(\frac{3}{8} -\frac{5}{4} \nu\right) v^{4}\;, \nonumber \\
    {}^{(2.5)}\mathbf{\boldsymbol\ell}_{1} &= (n\Sigma ) \, \mathbf{n} \left[-3 \nu\frac{\delta m}{m} (nv)^2+\frac{\delta m}{m}\left(\frac{1}{2} + \frac{3}{2} \nu\right) v^{2}\right] + (\Sigma v) \, \mathbf{n} \frac{\delta m}{m}\left(-\frac{1}{2} -\frac{13}{4} \nu\right) (nv) \nonumber \\
& + (nS) \, \mathbf{n} \left[-\frac{9}{2} \nu (nv)^2+\left(\frac{7}{2} -\frac{1}{2} \nu\right) v^{2}\right] + (Sv) \, \mathbf{n} \left(-\frac{3}{2} -\frac{7}{2} \nu\right) (nv) \nonumber \\
& + \frac{7}{4} \nu\frac{\delta m}{m} (nv) (n\Sigma ) \, \mathbf{v} + (\Sigma v) \, \mathbf{v} \frac{\delta m}{m}\left(-\frac{3}{2} + \frac{1}{2} \nu\right) + (nS) \, \mathbf{v} \left(-3 + 6 \nu\right) (nv) \nonumber \\
& + (Sv) \, \mathbf{v} \left(\frac{1}{2} -\frac{1}{2} \nu\right) + \, \mathbf{S} \left[\left(2 + \frac{5}{2} \nu\right) (nv)^2+\left(-\frac{3}{2} + \frac{1}{2} \nu\right) v^{2}\right] \nonumber \\
& + \, \mathbf{\Sigma} \left[\frac{7}{2} \nu\frac{\delta m}{m} (nv)^2+\frac{\delta m}{m}\left(\frac{3}{2} - \nu\right) v^{2}\right] \;, \nonumber \\
    {}^{(2.5)}\mathbf{\boldsymbol\ell}_{2} &= (n\Sigma ) \, \mathbf{n} \frac{\delta m}{m}\left(-\frac{1}{2} -\frac{3}{2} \nu\right) + (nS) \, \mathbf{n} \left(-\frac{1}{2} -2 \nu\right) + \, \mathbf{S} \left(\frac{1}{2} + 2 \nu\right) \nonumber \\
& + \, \mathbf{\Sigma} \frac{\delta m}{m}\left(\frac{1}{2} + \frac{3}{2} \nu\right)\;.
\end{align}}\noindent
\end{subequations}
As in the case for the energy, the next-to-leading term does not have
the same expression as the one published for instance in
Refs.~\cite{Kidder1995,Faye2006}, due to our different choice of spin
variables; but now the difference also shows up in the leading-order
spin contributions to $\mathbf{L}$ since it is actually the
next-to-leading order correction in the total angular momentum
$\mathbf{J}$. Finally, at next-to-next-to-leading order we have
\begin{subequations}
\begin{align}
\mathop{\boldsymbol\ell}_{S}{}_{\!\mathrm{3.5PN}}=
{}^{(3.5)}\mathbf{\boldsymbol\ell}_0
+{}^{(3.5)}\mathbf{\boldsymbol\ell}_1 \frac{G m}{r}
+{}^{(3.5)}\mathbf{\boldsymbol\ell}_2 \frac{G^2 m^2}{r^2}
+{}^{(3.5)}\mathbf{\boldsymbol\ell}_3 \frac{G^3 m^3}{r^3} \, ,
\end{align}
\allowdisplaybreaks{
\begin{align}
   {}^{(3.5)}\mathbf{\boldsymbol\ell}_{0} &= (\Sigma v) \, \mathbf{v} \frac{\delta m}{m}\left(-\frac{5}{16} + \frac{19}{8} \nu -\frac{75}{16} \nu^2\right) v^{4} + (Sv) \, \mathbf{v} \left(-\frac{5}{16} + \frac{35}{16} \nu -\frac{65}{16} \nu^2\right) v^{4} \nonumber \\
& + \, \mathbf{S} \left(\frac{5}{16} -\frac{35}{16} \nu + \frac{65}{16} \nu^2\right) v^{6} + \, \mathbf{\Sigma} \frac{\delta m}{m}\left(\frac{5}{16} -\frac{19}{8} \nu + \frac{75}{16} \nu^2\right) v^{6} \;, \nonumber \\ 
    {}^{(3.5)}\mathbf{\boldsymbol\ell}_{1} &= (n\Sigma ) \, \mathbf{n} \left[\frac{\delta m}{m}\left(\frac{15}{4} \nu -\frac{75}{8} \nu^2\right) (nv)^4+\frac{\delta m}{m}\left(-\frac{15}{2} \nu + 9 \nu^2\right) (nv)^2 v^{2} \right. \nonumber \\
& \quad \left. +\frac{\delta m}{m}\left(\frac{3}{8} + \frac{27}{8} \nu -\frac{77}{8} \nu^2\right) v^{4}\right] + (\Sigma v) \, \mathbf{n} \left[\frac{\delta m}{m}\left(\frac{51}{16} \nu + \frac{15}{8} \nu^2\right) (nv)^3 \right. \nonumber \\
& \quad \left. + \frac{\delta m}{m}\left(-\frac{3}{8} -\frac{23}{4} \nu + \frac{243}{16} \nu^2\right) (nv) v^{2}\right]  + (nS) \, \mathbf{n} \left[\left(\frac{45}{8} \nu -\frac{135}{8} \nu^2\right) (nv)^4 \right. \nonumber \\
& \quad \left. +\left(-\frac{33}{4} \nu + \frac{33}{4} \nu^2\right) (nv)^2 v^{2}+\left(\frac{33}{8} -\frac{83}{8} \nu -\frac{11}{8} \nu^2\right) v^{4}\right] \nonumber \\
& + (Sv) \, \mathbf{n} \left[\left(-\frac{3}{4} + \frac{9}{2} \nu + \frac{33}{4} \nu^2\right) (nv)^3+\left(-\frac{17}{8} + \frac{5}{16} \nu + \frac{109}{8} \nu^2\right) (nv) v^{2}\right] \nonumber \\
& + (n\Sigma ) \, \mathbf{v} \left[\frac{\delta m}{m}\left(-\frac{21}{16} \nu + \frac{93}{8} \nu^2\right) (nv)^3+\frac{\delta m}{m}\left(\frac{1}{4} \nu -\frac{31}{16} \nu^2\right) (nv) v^{2}\right] \nonumber \\
& + (\Sigma v) \, \mathbf{v} \left[\frac{\delta m}{m}\left(-\frac{3}{4} \nu -\frac{19}{8} \nu^2\right) (nv)^2+\frac{\delta m}{m}\left(-\frac{21}{8} + \frac{59}{8} \nu -\frac{35}{8} \nu^2\right) v^{2}\right] \nonumber \\
& + (nS) \, \mathbf{v} \left[\left(\frac{3}{4} -\frac{15}{2} \nu + 27 \nu^2\right) (nv)^3+\left(-\frac{17}{4} + \frac{295}{16} \nu -\frac{75}{4} \nu^2\right) (nv) v^{2}\right] \nonumber \\
& + (Sv) \, \mathbf{v} \left[\left(\frac{1}{8} \nu -\frac{43}{4} \nu^2\right) (nv)^2+\left(\frac{3}{8} -\frac{31}{8} \nu + \frac{5}{8} \nu^2\right) v^{2}\right] \nonumber \\
& + \, \mathbf{S} \left[\left(-\frac{21}{8} \nu -\frac{33}{8} \nu^2\right) (nv)^4+\left(3 -\frac{9}{4} \nu -\frac{21}{4} \nu^2\right) (nv)^2 v^{2}+\left(-\frac{9}{8} + \frac{47}{8} \nu -\frac{5}{8} \nu^2\right) v^{4}\right] \nonumber \\
& + \, \mathbf{\Sigma} \left[\frac{\delta m}{m}\left(-\frac{33}{8} \nu + \frac{9}{8} \nu^2\right) (nv)^4+\frac{\delta m}{m}\left(\frac{39}{4} \nu -17 \nu^2\right) (nv)^2 v^{2} \right. \nonumber \\
& \quad \left. +\frac{\delta m}{m}\left(\frac{21}{8} -\frac{33}{4} \nu + \frac{47}{8} \nu^2\right) v^{4}\right] \;,\nonumber \\
    {}^{(3.5)}\mathbf{\boldsymbol\ell}_{2} &= (n\Sigma ) \, \mathbf{n} \left[\frac{\delta m}{m}\left(-\frac{469}{8} \nu -\frac{51}{4} \nu^2\right) (nv)^2+\frac{\delta m}{m}\left(\frac{7}{4} -\frac{149}{4} \nu + \frac{33}{4} \nu^2\right) v^{2}\right] \nonumber \\
& + (nS) \, \mathbf{n} \left[\left(-\frac{973}{8} \nu -\frac{39}{2} \nu^2\right) (nv)^2+\left(\frac{55}{4} -\frac{311}{4} \nu + 6 \nu^2\right) v^{2}\right] \nonumber \\
& + (\Sigma v) \, \mathbf{n} \frac{\delta m}{m}\left(-\frac{17}{8} + \frac{103}{4} \nu -\frac{25}{2} \nu^2\right) (nv) + (Sv) \, \mathbf{n} \left(-\frac{29}{8} + \frac{247}{4} \nu -\frac{31}{2} \nu^2\right) (nv) \nonumber \\
& + (n\Sigma ) \, \mathbf{v} \frac{\delta m}{m}\left(-\frac{1}{8} + \frac{177}{2} \nu + 3 \nu^2\right) (nv) + (\Sigma v) \, \mathbf{v} \frac{\delta m}{m}\left(-\frac{7}{4} -\frac{257}{8} \nu + \frac{21}{4} \nu^2\right) \nonumber \\
& + (nS) \, \mathbf{v} \left(-\frac{117}{8} + \frac{731}{4} \nu + \frac{27}{2} \nu^2\right) (nv) + (Sv) \, \mathbf{v} \left(\frac{25}{4} -\frac{571}{8} \nu + \frac{9}{2} \nu^2\right) \nonumber \\
& + \, \mathbf{S} \left[\left(\frac{17}{2} -\frac{163}{8} \nu + \frac{35}{2} \nu^2\right) (nv)^2+\left(-\frac{41}{4} + \frac{373}{8} \nu -\frac{13}{2} \nu^2\right) v^{2}\right] \nonumber \\
& + \, \mathbf{\Sigma} \left[\frac{\delta m}{m}\left(\frac{1}{2} -\frac{43}{8} \nu + \frac{33}{2} \nu^2\right) (nv)^2+\frac{\delta m}{m}\left(\frac{7}{4} + \frac{153}{8} \nu -\frac{31}{4} \nu^2\right) v^{2}\right] \;,\nonumber \\ 
    {}^{(3.5)}\mathbf{\boldsymbol\ell}_{3} &= (n\Sigma ) \, \mathbf{n} \frac{\delta m}{m}\left(\frac{1}{2} -2 \nu -\frac{3}{2} \nu^2\right) + (nS) \, \mathbf{n} \left(\frac{3}{2} -\frac{13}{4} \nu -2 \nu^2\right) + \, \mathbf{S} \left(-\frac{3}{2} + \frac{13}{4} \nu + 2 \nu^2\right) \nonumber \\
& + \, \mathbf{\Sigma} \frac{\delta m}{m}\left(-\frac{1}{2} + 2 \nu + \frac{3}{2} \nu^2\right) \;.
\end{align}}
\end{subequations} 

\section{Gauging away the 3PN spin-orbit terms}
\label{3PNgauge}

The metric regularized at the location of the two particles, as was
presented in Section~\ref{subsec:regularizedmetricCM}, displays terms
which are formally of 3PN order, that is to say of order
$\mathcal{O}(1/c^{8})$ in $g_{00}$ and $\mathcal{O}(1/c^{7})$ in
$g_{0i}$. These terms are given by the expressions \eqref{g00S1CM3PN}
and \eqref{g0iS1CM3PN}:
\begin{align} \label{eq:gaugetermsing}
	\left(\mathop{g}_{S}{}^{\!\mathrm{3PN}}_{00}\right)_1=&
\frac{G^{2}m \nu}{c^{8}r^{3}}\bigg\{(n,\Sigma,v)\left(2 -10\frac{\delta m}{m}\right) (nv)- 24 (nv)(n,S,v)\bigg\},\\
	\left(\mathop{g}_{S}{}^{\!\mathrm{3PN}}_{0i}\right)_1=&
\frac{G^{2}m\nu}{c^{7}r^{3}}\bigg\{2 (nv) (\mathbf{n} \times \mathbf{\Sigma} )^{i} - \frac{2}{3}(\mathbf{v} \times \mathbf{\Sigma})^{i}\bigg\}.
\end{align}

In this Appendix, we show that they are actually a pure coordinate
effect, in the sense that they can be eliminated by a gauge
transformation. We already encountered such a gauge transformation in Ref.~\cite{Blanchet2011} and in
Paper I (see the footnote in Section~VI~B there), where it was used to
remove the $\mathcal{O}(1/c^{6})$ terms appearing in the
acceleration. Here we generalize this transformation to cancel as well
the 3PN terms in the components of the regularized metric. Notice
that, for $g_{00}$ and $g_{0i}$, we do not control the metric
components in the bulk at the corresponding order. The transformation
we present here is only suitable for removing these terms in the
regularized metric.

We consider a coordinate transformation $x'^{\mu} = x^{\mu} +
\xi^{\mu}$, which has to satisfy $\Box \xi^{\mu} = 0$ (at dominant
order) to respect the harmonic gauge condition. We will need to have
$\xi^{0}=\mathcal{O}(1/c^{7})$ and
$\xi^{i}=\mathcal{O}(1/c^{6})$. Since the components $g_{ij}$ of the
metric must remain unaffected in the bulk at order
$\mathcal{O}(1/c^{6})$ (see \eqref{gijSCM}), we further require that
the component $\xi^{i}$ is a mere function of time, \textit{i.e.} does
not depend on the field point $\mathbf{x}$. We allow a spatial
dependence in the component $\xi^{0}$, but the harmonicity condition
$\Delta \xi^{0} = \mathcal{O}(1/c^{9})$ constrains its
structure. Hence, we take the following Ansatz for the transformation:
\begin{subequations}\label{eq:gaugeAnsatz}
\begin{align}
  \xi^{0} &= \frac{Gm^{2}\nu}{c^{7}}\left[ A + B_{1}\frac{1}{r_{1}} + B_{2}\frac{1}{r_{2}} + C_{1}^{i}\partial_{i}\left(\frac{1}{r_{1}}\right) + C_{2}^{i}\partial_{i}\left(\frac{1}{r_{2}}\right)\right] \;, \\
  \xi^{i} &= \frac{G^{2}m\nu}{c^{6}} D^{i} \; ,
\end{align}
\end{subequations}
where $A$, $B_{1}$, $B_{2}$, $C_{1}^{i}$, $C_{2}^{i}$ and $D^{i}$ are
all mere functions of time. This gauge transformation has to be
symmetric by exchange of the two bodies, which relates $B_{1}$ to
$B_{2}$, $C_{1}^{i}$ to $C_{2}^{i}$ and constrains the expression of
$A$ and $D^{i}$. We could have considered higher order derivatives of
$1/r$ in $\xi^{0}$, but we found that the above structure was
sufficient for our purpose.

With our simple choice for $\xi^{i}$, the gauge transformation changes the metric components according to,
 (see Eqs~(6.9) in
\cite{Blanchet2001a}):
\begin{subequations} \label{eq:gaugetrans}
  \begin{align}
    \delta_{\xi}g_{00} &= 2 \partial_{0}\xi^{0} + \mathcal{O}(1/c^{10}) \; , \\
    \delta_{\xi}g_{0i} &= - \partial_{0}\xi^{i} + \partial_{i}\xi^{0} + \mathcal{O}(1/c^{9}) \; , \\
    \delta_{\xi}g_{ij} &= - \partial_{i}\xi^{j} - \partial_{j}\xi^{i}
    + \mathcal{O}(1/c^{8}) \; ,
\end{align}
\end{subequations}
while the acceleration transforms as (see Eqs.~(4.8) in
\cite{Arun2008} for more general expressions):
\begin{equation}
  \delta_{\xi}a_{1}^{i} = \frac{\ud^{2}}{\ud t^{2}} \xi^{i}(t) + \mathcal{O}(1/c^{8}) \; .
\end{equation}

This shows that $\xi^{i}$ is the same as we used in Paper I, and which
was called $\delta X^{i}$ there. Solving for the other unknown
functions of time by applying \eqref{eq:gaugetrans} to the structure
\eqref{eq:gaugeAnsatz}, we find the following expressions:
\begin{subequations}
  \begin{align} \label{eq:gaugeres}
    A &= \frac{1}{r^{2}} \left[ 4 (\mathbf{n},\mathbf{v},\mathbf{S}) + \frac{3}{2} \frac{\delta m}{m} (\mathbf{n},\mathbf{v},\mathbf{\Sigma}) \right] \; , \\
    B_{2} &= - B_{1} = -\frac{2}{3 r} (\mathbf{n},\mathbf{v},\mathbf{\Sigma}) \; , \\
    \mathbf{C}_{2} &= \mathbf{C}_{1} = -\frac{5}{6} (\mathbf{n},\mathbf{v},\mathbf{\Sigma}) \mathbf{n} + \frac{1}{3} \mathbf{v}\times\mathbf{\Sigma} - (nv) \mathbf{n}\times\mathbf{\Sigma} \; , \\
    \mathbf{D} &= -\frac{1}{r^{2}} \mathbf{n}\times\mathbf{\Sigma} \;
    .
\end{align}
\end{subequations}
This solution allows us, within the class of harmonic coordinates, to
cancel the 3PN terms \eqref{eq:gaugetermsing} in the regularized
metric, which are therefore pure gauge.

\bibliography{biblioBMFB}

\end{document}